 \documentclass[preprint,12pt,authoryear]{elsarticle}

\usepackage{amssymb}

\usepackage{lineno}
\usepackage[normalem]{ulem}
\usepackage{mdframed}
\usepackage{xcolor}
\usepackage{multirow}
\usepackage{amsmath}

\journal{Icarus}

\def\deg{\ifmmode^\circ\else$\null^\circ$\fi}
\def\N2{N$_2$}
\def\um{$\mu$m }

\def\icarus{\ref@jnl{Icarus}}

\begin{document}

\begin{frontmatter}

\title{Modeling Pluto's Minimum Pressure:  Implications for Haze Production}


\author[1]{Perianne E. Johnson\footnote{Corresponding author}}
\author[2]{Leslie A. Young}
\author[2]{Silvia Protopapa}
\author[3]{Bernard Schmitt}
\author[3]{Leila R. Gabasova}
\author[4]{Briley L. Lewis}
\author[5]{John A. Stansberry}
\author[6]{Kathy E. Mandt}
\author[7,8]{Oliver L. White}

\address[1]{University of Colorado, Boulder, CO, 80303, United States}
\address[2]{Southwest Research Institute, Boulder, CO, 80302, United States}
\address[3]{Universit\'e Grenoble Alpes, CNES, IPAG, Grenoble Cedex 9, France}
\address[4]{University of California, Los Angeles, CA, 90095, United States}
\address[5]{Space Telescope Science Institute, Baltimore, MD, 21218, United States}
\address[6]{Johns Hopkins University Applied Physics Laboratory, Laurel, MD, 20723, United States}
\address[7]{NASA Ames Research Center, Moffett Field, CA, 94035, United States}
\address[8]{SETI Institute, Mountain View, CA, 94043, United States}

\date{20 August 2019}


\begin{abstract}
Pluto has a heterogeneous surface, despite a global haze deposition rate of $\sim$1 $\mu$m per orbit \citep{Cheng+2017, Grundy+2018}. While there could be spatial variation in the deposition rate, this has not yet been rigorously quantified, and naively the haze should coat the surface more uniformly than was observed. One way (among many) to explain this contradiction is for atmospheric pressure at the surface to drop low enough to interrupt haze production and stop the deposition of particles onto part of the surface, driving heterogeneity. If the surface pressure drops to less than 10$^{-3}$ - 10$^{-4}$ $\mu$bar and the CH$_4$ mixing ratio remains nearly constant at the observed 2015 value, the atmosphere becomes transparent to ultraviolet radiation \citep{Young+2018}, which would shut off haze production at its source. If the surface pressure falls below 0.06 $\mu$bar, the atmosphere ceases to be global, and instead is localized over only the warmest part of the surface, restricting the location of deposition \citep{Spencer+1997}. In Pluto's current atmosphere, haze monomers collect together into aggregate particles at beginning at 0.5 $\mu$bar; if the surface pressure falls below this limit, the appearance of particles deposited at different times of year and in different locations could be different. We use VT3D, an energy balance model \citep{Young2017}, to model the surface pressure on Pluto in current and past orbital configurations for four possible static \N2 ice distributions: the observed northern hemisphere distribution with (1) a bare southern hemisphere, (2) a south polar cap, (3) a southern zonal band, and finally (4) a distribution that is bare everywhere except inside the boundary of Sputnik Planitia. We also present a sensitivity study showing the effect of mobile \N2 ice. By comparing the minima of the modeled pressures to the three haze-disruption pressures, we can determine if or when haze production is disrupted. We find that Pluto's minimum surface pressure in its current orbit is predicted to be between 0.01 - 3 $\mu$bar, and that over the past 10 million years the surface pressure has not fallen below 0.004 $\mu$bar. According to our model, southern \N2 ice is required for haze aggregation to be interrupted, and southern \N2 with very low thermal inertia is required for the possibility of a local atmosphere. 
\end{abstract}

\begin{keyword}
	Pluto;  Pluto, atmosphere; Pluto, surface; Ices
	
	
	
\end{keyword}

\end{frontmatter}

\section{Introduction}
The New Horizons mission to Pluto revealed a surprisingly active surface, with dramatic albedo, color, and composition contrasts \citep{Stern+2015}. The flyby also detected haze in the atmosphere, and haze particles should settle through the atmosphere and be deposited onto the surface. These two observations presented a major question: how is the heterogeneity maintained despite a global blanket of deposited haze particles on the surface? This work investigates one possible answer to this question, which is that the atmospheric pressure could drop low enough for long enough over Pluto's orbit to disrupt haze production at its source, preventing the haze particles from being deposited onto the surface.

Pluto's normal reflectance varies across its surface by over a factor of ten, with some regions reaching a normal reflectance value of unity and the darkest regions dropping to a minimum of 0.08 \citep{Buratti+2017}. The equatorial region is dark and red, interrupted by bright, more neutral Sputnik Planitia (the expansive volatile-ice sheet that makes up the western half of Tombaugh Regio, Pluto's ``heart''; hereafter called SP); midlatitudes, especially where covered by volatile ices, are similar to SP's neutral color, while the north polar region (north of 60$^\circ$N) has a yellow hue \citep{Stern+2015, Olkin+2017, Protopapa+2020}. Composition also varies across the encounter hemisphere, with SP showing very strong \N2 and CH$_4$ spectral signatures, while the dark equatorial region appears to be free of both species and instead has a spectrum that is consistent with tholins, an unknown mix of hydrocarbons and carbonaceous material produced by energetic radiation (including cosmic rays and UV) interactions with \N2 and CH$_4$ \citep{Protopapa+2017, Protopapa+2020, Schmitt+2017}.

New Horizons observed haze extending up to 300 km, which was globally present but not spatially uniform \citep{Stern+2015, Cheng+2017}. The haze was brighter towards northern latitudes rather than in the direction of the Sun. \citet{Cheng+2017} compared I/F values for 44\deg N and -0.5\deg S, and found that the northern latitude haze was systematically brighter by factors of 2 to 3 compared with the equatorial haze. Pluto's haze is created by radiolysis and photolysis of the atmospheric species, primarily CH$_4$, \N2, and CO, using a variety of energy sources \citep{Mandt+2020}. Solar UV radiation, including Lyman-$\alpha$, is important to haze photochemistry and its flux is greatest at the sub-solar point on Pluto, which could potentially increase the haze production rate there. Lyman-$\alpha$ is also scattered by the interplanetary medium and impinges on Pluto's nightside. Cosmic rays, another important energy source, will hit Pluto's atmosphere isotropically. Solar wind particle fluxes were time-variable, as measured by New Horizons, and their interaction with Pluto's atmosphere is uncertain \citep{Bagenal+2016}. Accounting for the space- and time-variability of these energy sources for haze production makes it difficult to predict the expected variability in haze deposition and its distribution on the surface. 

Deposition rates from \citet{Cheng+2017}, \citet{Grundy+2018}, and \citet{Krasnopolsky2020} all predict that a layer of haze particles roughly one micron thick would accumulate over one Pluto orbit, amounting to more than 10 m over the age of the solar system. While the haze was not observed to be spatially uniform and the production mechanism might vary spatially or temporally as well, these authors do not address such variations and instead present deposition rates as approximate global averages. \citet{Grundy+2018} conclude that in order to produce the observed heterogeneity, the haze particles must either be differentially processed once on the surface, or the production and deposition must be spatially variable, although they do not suggest a mechanism to cause this spatial variability. \citet{Cheng+2017} suggest that haze deposition may be interrupted by atmospheric collapse (here, we define collapse to mean when the atmosphere is localized and ``patchy'' rather than global). \citet{Grundy+2018} discuss this possibility as well, and also raise other mechanisms such as a spatially or temporally variable gaseous CH$_4$ column, or the movement of haze particles by wind once they have settled on the surface (explored further in \citet{Forget+2017} and \citet{Bertrand+2019b}). Neither \citet{Cheng+2017} nor \citet{Grundy+2018} quantifies the possibility of interrupting or diminishing haze production within the atmosphere. \citet{Bertrand+2017} explored the production and atmospheric transport of haze particles, and found that the column mass of haze aerosols in the atmosphere varies spatially by a factor of 10 if there is no condensation of volatiles at the southern pole in winter, or only a factor of 2 if south pole condensation is allowed. \citet{Forget+2017} used a Pluto Global Climate Model to investigate the atmospheric circulation, and found that zonal flows were on the order of a few m s$^{-1}$ (varying with latitude and with the assumed \N2 distribution), while meridional flows were much smaller, on the order of a few tens of cm s$^{-1}$. Both of these flows could redistribute the haze particles, either through atmospheric transport pre-deposition, or near-surface winds could blow haze particles around and collect them into localized regions post-deposition.

If the atmospheric pressure at the surface gets low enough, haze production may be altered, suppressed or stopped completely. \citet{Gao+2017} use a microphysical model to show that haze particles begin to grow at around 150 to 300 km altitude (depending on the size of the initial monomers) and this growth continues as the particles fall to the surface. This altitude range encompasses pressures from ~0.1 $\mu$bar to 0.6 $\mu$bar; we select 0.5 $\mu$bar to be representative of the pressure level where aggregation begins. Below the 150-300 km level, sedimentation timescales are longer than coagulation timescales, due to the increased atmospheric density, allowing haze particles more time to collide and coagulate into larger aggregates \citep{Cheng+2017, Gao+2017}. If the surface pressure drops below this level, monomer haze particles (which are created around 1000 km altitude in the 2015 atmosphere) may sediment out onto the surface instead of aggregates, potentially changing the appearance on the surface. We refer to this as a ``non-aggregating'' atmosphere. This pressure limit is based on the atmosphere as observed by New Horizons in 2015; we assume that the pressure level where haze aggregation occurs stays constant throughout Pluto's orbit. However, the sedimentation and coagulation timescales depend on quantities such as the atmospheric density, temperature, and dynamic viscosity and the sizes of the monomer and aggregate particles, and the temporal behavior of these quantities has not yet been well studied for Pluto. For surface pressures less than $\sim$0.06 $\mu$bar Pluto cannot support a global atmosphere \citep{Spencer+1997}, and instead the atmosphere becomes local, or patchy, which would restrict the region in which haze particles are deposited. Additionally, if the surface pressure drops to less than 10$^{-3}$ - 10$^{-4}$ $\mu$bar, the atmosphere would be transparent to ultraviolet radiation \citep{Young+2018}. This would shut off the photolysis of atmospheric \N2 and CH$_4$, suppressing haze production at its source \citep{Gao+2017}, while simultaneously boosting the photolysis of surface ices and existing tholins, which can lead to a different composition and appearance of tholins than those produced in the atmosphere \citep{Bertrand+2019}.    

Pluto's obliquity varies with a 2.8-million year period, and this obliquity cycle creates “extreme seasons” during which perihelion occurs simultaneously with northern summer solstice (most recently occurred 0.9 Mya) or aphelion occurs simultaneously with northern summer solstice (most recently occurred 2.4 My ago) \citep{Earle+2017, Bertrand+2018}. During these two extreme orbital configurations the minimum surface pressure will be different from that in the current configuration, providing an opportunity for historic haze disruption that might not be seen in today's Pluto. This could affect the present-day surface heterogeneity. 

\citet{TraftonStern1983} considered a CH$_4$ atmosphere (CH$_4$ was then the only species detected at Pluto) and predicted a globally-uniform surface pressure for CH$_4$ column abundances greater than 6.7 cm-Am (using the now-known surface gravity of 0.62 ms$^{-2}$, this corresponds to a pressure of 0.3 $\mu$bar). At the time, the best estimate for the column abundance was 27 $\pm$ 7 m-Am (12 $\pm$ 3 $\mu$bar), which implied that energy could be efficiently transported from high-insolation areas to low-insolation areas, and that vapor pressure equilibrium could maintain a uniform surface temperature of 58 $\pm$ 0.9 K. After the discovery of Pluto's atmosphere via occultations in 1988 \citep{Elliot+1989, Hubbard+1988}, and the detection of abundant \N2 by \citet{Owen+1993}, \citet{HansenPaige1996} adapted their existing Triton energy balance model to Pluto. They found that volatile transport would be a significant process coupling the surface and atmosphere, allowing surface ices to move around on seasonal timescales. They also found that perennial zonal bands of ice could form in their model, as opposed to perennial polar ice caps, due to Pluto's high obliquity. For some cases, ``polar bald spots'' were created by sublimation that began at the center of a polar cap rather than at the equatorward edge. \N2 ice temperatures between 30 and 40 K were predicted, based on the balance between insolation, infrared thermal emission, conduction to and from the subsurface, and the latent heat of subliming and condensing \N2.

\citet{Bertrand2016} used a simplified Pluto GCM to simulate Pluto's climate and volatile transport for thousands of orbits in a reasonable computation time. They found that, for an initial globally uniform distribution of \N2 ice and thermal inertias above 700 tiu (Thermal Inertia Units, J m$^{-2}$ K$^{-1}$ s$^{-1/2}$), all of the \N2 ice migrated into their modeled 3-km deep SP basin within 10,000 Earth years. This motivated the ``strawman'' example we present in Section 3.2 using a SP-only \N2 distribution. For lower thermal inertias, their model had seasonal deposits of \N2 ice outside of SP. When the thermal inertia of the \N2 ice was $>$700 tiu, their model predicted pressures that were consistent with pre-existing occultation measurements (implying a roughly two- to three-fold increase in pressure between 1988 and 2015), as well as a peak value of about 11.5 $\mu$bar near 2015. \citet{Bertrand+2018} explored the \N2 cycles using their parameterized Pluto GCM on million year timescales, capturing the response to the obliquity cycles described above. They found that a net value of 1 km of \N2 ice has sublimed from the northern edge of SP and recondensed onto the southern edge over the past 2 million years, driven by the change in subsolar latitude at perihelion, which shifted from the southern hemisphere to the north (and is now moving back towards the south, currently near 0$^{\deg}$, see Figure 1 in \citet{Earle+2017}). They also found that over millions of years the surface pressure on Pluto never drops below tens of nanobars, nor exceeds tens of microbars. 

We aim to test the hypothesis of haze disruption via thermal modelling of the surface. Our model, VT3D, is described in Section 2, along with our choices for thermal parameters and the distribution of surface volatiles. Sections 3 presents the resulting pressure evolution curves for the current Pluto orbit and past orbits with different obliquities and subsolar latitudes at perihelion, assuming four different \N2 distributions, as well as a sensitivity study for the effect of mobile \N2 ice. Finally, we discuss the implications of these modelled pressure curves in relation to haze production in Section 4.

\section{Methods}

\subsection{VT3D Model Overview}
This section provides an overview of the Volatile Transport Three Dimensional (VT3D) model as used in this study; for a complete description of the model and its full capabilities, see \citet{Young2017}. VT3D is an energy balance model, including thermal conduction into and within a substrate, internal heat (not used here), latent heat of sublimation, insolation, and thermal emission. Locally, the energy balance equation is:
\begin{equation}
\frac{S_{1AU}(1-A)\mu}{r^2} - \epsilon\sigma T^4 - k\frac{dT}{dz} + L\dot{m} = 0
\end{equation}
where $S_{1AU}$ is the solar constant (1361 W m$^{-2}$), $A$ is the Bond albedo of the surface, $\mu$ is the solar incidence angle at the given location, $r$ is the heliocentric distance in AU, $\epsilon$ is the emissivity of the surface, $\sigma$ is the Stefan-Boltzmann constant, $T$ is the volatile temperature, $k$ is the thermal conductivity, $z$ is the depth beneath the surface (zero at the surface and decreasing downward), $L$ is the latent heat of sublimation, and $\dot{m}$ is the condensation rate, in mass per area per time. The partition between sublimation and conduction is determined by global mass balance \citep{Young2012, Young2013}, since the rate of change of the total atmospheric bulk (areal integral of $\dot{m}$) is related to the change in \N2 ice temperature through the change in the surface pressure and atmospheric column density. As implemented here, VT3D depends on three free parameters (the Bond albedo, \textit{A}, the emissivity, $\epsilon$, and the seasonal thermal inertia, $\Gamma$, of the surface \N2 ice) as well as on the spatial distribution of \N2 ice. \N2 is the dominant atmospheric constituent and it is more volatile than  the minor constituents of CH$_4$ and CO, so we consider only the \N2 temperature when we model the atmospheric pressure. 

We run VT3D using the explicit form of the equations (rather than its semi-implicit Crank-Nicholson scheme). The explicit scheme is only stable for small timesteps; we calculate the temperature at 500 points per Pluto orbit, corresponding to a timestep of about 0.5 Earth year. The volatiles are discretized vertically into \textit{J} = 40 layers for a total depth of roughly 10 thermal skin depths. The temperature at the next timestep of a given layer depends on the temperature at the current timestep in the layer above, in the layer itself, and in the layer below. To evaluate the insolation term, we average the insolation at the start and end of the current timestep: $(S_n+S_{n+1})/2$, where subscript \textit{n} represents the current timestep, and \textit{n}+1 is the next timestep. We use the diurnally- and spatially-averaged insolation, as discussed more in the following section. To evaluate the average thermal emission term for the timestep from $t_n$ to $t_{n+1}$, VT3D uses the first-order Taylor expansion of $T^4$: 
$\epsilon\sigma T_{0,n}^4 + 2\epsilon\sigma T_{0,n}^3(T_{0,n+1} - T_{0,n})$, where the first subscript indicates the layer (0 corresponds to the top layer) and the second indicates the timestep. The conduction term is discretized using a first-order finite difference scheme; for example the term describing conducted heat from the layer below into the top layer is: $\sqrt{\omega}\Gamma(T_{0,n}-T_{1,n})/\delta$, where $\omega$ is the orbital frequency of Pluto, in seconds, and $\delta$ is the dimensionless distance between layers. The sublimation rate is related to the rate of change of temperature since we assume vapor pressure equilibrium; in response to an increase in the ice temperature, the vapor pressure above it must also increase, which means particles sublime from the ice surface, removing latent heat. Thus, the sublimation term is written: $\Phi^A(T_{0,n+1} - T_{0,n})$, where $\Phi^A$ is given by: $\Phi^A = L^2m_{N_2}p\omega/(f_vgk_BT_{0,n}^2\tau)$ ($L$ is the latent heat of sublimation for \N2, $m_{N_2}$ is the molecular mass of \N2, $p$ is the vapor pressure at temperature $T_{0,n}$, $f_v$ is the fraction of the surface covered by \N2, $g$ is the surface gravity, $k_B$ is the Boltzmann constant, and $\tau$ is a dimensionless time step). After inserting these terms into Equation 1, temperatures at the next timestep are a function of temperatures at the current timestep and various parameters of the \N2 ice. VT3D finds the temperatures by stepping forward in time for one Pluto orbit.

VT3D begins with an analytic approximation to the solution, which is used as the initial guess in the more accurate numerical solution to decrease convergence time. A description on how to implement the analytic solution for quick calculation is included in the Appendix. 

To convert temperatures into pressures, we use the equation for solid \N2 vapor pressure as a function of temperature presented in \citet{FraySchmitt+2009}:
\begin{equation}
ln\left(P_{sub}\right) = A_0 + \sum_{i=1}^n\frac{A_i}{T^i}
\end{equation}
\citet{FraySchmitt+2009} compile previously-published empirical relations and experimental data to find the best-fit coefficients $A_i$ for solid \N2 ice, with separate sets of coefficients for the $\alpha$- and $\beta$-crystalline phases, shown in Table \ref{table:fs09}.

\begin{table}[]
	\caption{Coefficients needed to calculate the equilibrium vapor pressure as a function of temperature.}
	\begin{center}
	\footnotesize
	\begin{tabular}{|c|c|c|c|c|c|c|}
		\hline
		& $A_0$   & $A_1$    & $A_2$    & $A_3$    & $A_4$      & $A_5$     \\ \hline
		\hline
		$\alpha$-phase & 12.404174  & -807.35728 & -3925.5143  & 62965.429  & -463269.99  & 1.324999.3 \\ \hline
		$\beta$-phase  & 8.51384232 & -458.386541 & -19871.6407 & 480001.675 & -4523786.13& 0       \\ \hline
	\end{tabular}
		
	\label{table:fs09}
	\end{center}
\end{table}

\subsection{Volatile Distribution}

\begin{figure}
	\begin{center}
		\includegraphics[width =\textwidth]{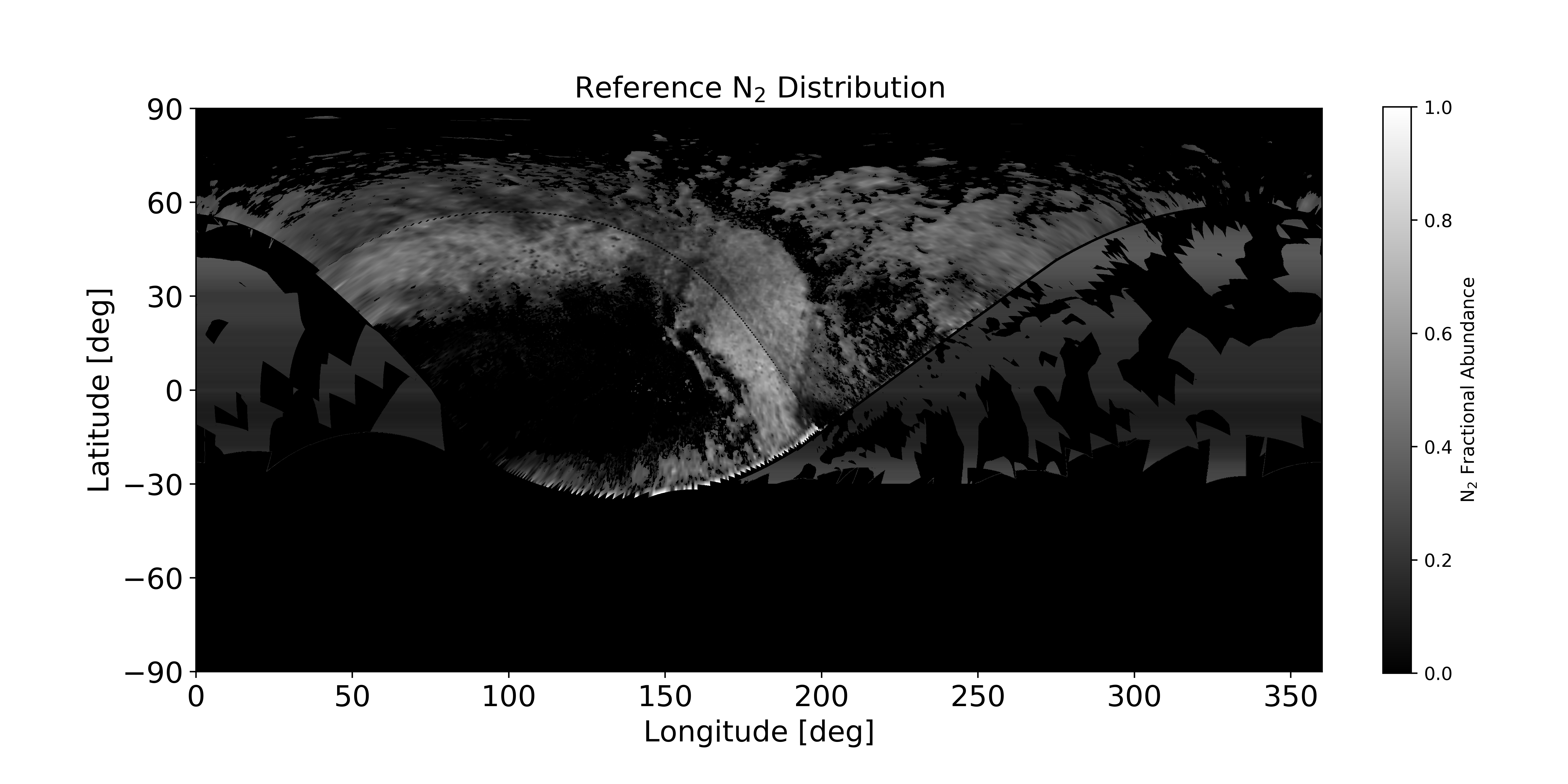}
		\caption{The spatial distribution of \N2 ice in our reference map. The grayscale represents the fractional abundance of \N2 ice coverage in that pixel. This distribution incorporates data from \citet{Protopapa+2017}, \citet{Schmitt+2017}, and Gabasova et al. (2020).}
		\label{fig:n2}
	\end{center}
\end{figure}

Observations of the surface volatile distribution were performed by the LEISA infrared spectrometer on the New Horizons spacecraft. \N2 ice is detectable by a weak 2.15 \um spectral feature, but only for sufficiently large grain sizes. Its presence can also be inferred from a wavelength shift in CH$_4$ spectral bands that occur when CH$_4$ is dissolved in \N2, and from the overall infrared brightness \citep{Protopapa+2017, Schmitt+2017}. \citet{Protopapa+2017} use a combination of these features along with Hapke radiative transfer modeling to produce a map of \N2 on the encounter hemisphere. Other analyses relying on spectral parameters like band depth or equivalent width are not able to distinguish between relative abundance changes and grain size changes across the surface. \citet{Protopapa+2017} produce separate fractional abundance and grain size maps. The modeled grain sizes (where grain size refers to distance between scattering centers, see \citet{Hapke}) range from a few centimeters to larger than 1 meter. The fractional abundance map highlights the large, flat ice sheet of SP, along with a latitudinal band stretching from 35\deg N to 55\deg N, as the main \N2 reservoirs on the surface, containing up to about 60\% \N2 (assuming an areal mixture with the other species that are present in that region). The fractional abundance of \N2 is the fraction of a given area that needs to be covered by \N2 to produce a spectra consistent with the observed spectra; the remaining area is covered by other species, namely CH$_4$, water ice, and tholins in \citet{Protopapa+2017}. \citet{Schmitt+2017} present a spatial distribution map of the \N2 ice band depth, as well as a map of the presence of the \N2-rich phase (called the `CH4 band position index' map) and their correlation, which make use of principal component analysis to reduce the noise and remove some instrument artifacts in the spectro-images of the high resolution LEISA data. \citet{Lewis+2019} created a \N2 presence map which combines the band depth map from \citet{Schmitt+2017} and the fractional abundance map from \citet{Protopapa+2017}. This map assumes a band depth above 0.005 or a fractional abundance of greater than 0\% indicates the presence of \N2. 

The high resolution LEISA images are limited to the encounter hemisphere, which was visible to the spacecraft during the flyby. The encounter hemisphere is centered near SP at longitudes around 150\deg, where the high resolution coverage reaches from north pole to 30\deg S. The region tapers off to the east and west until it reach the permanently lit north polar region extending out to 60\deg N. Much of the southern hemisphere (south of 40\deg S) is currently in polar night.

Gabasova et al. (2020) have used lower-resolution approach data in combination with the higher resolution flyby data to create a global \N2 distribution map that includes both the non-encounter and encounter hemispheres. This map shows the spatial distribution of the 2.15 $\mu$m \N2 band depth alone, and does not consider the shifting of the CH$_4$ bands nor the overall brightness of the pixel. A band depth value of 0.005 or greater indicates the presence of \N2 ice; however since band depth does not directly relate to the fractional abundance of \N2, this cannot be directly converted into a fractional abundance map. Attempts to correlate band depth and fractional abundance using the overlapping encounter hemisphere data did not yield a clear relationship, due in part to the grain size dependence of band depth. Instead, we turn the band depth map from Gabasova et al. (2020) into a \N2 presence map by applying a band depth threshold of 0.005, analogous to the procedure used by \citet{Lewis+2019}. We then find the zonal-average fractional abundance in each latitude band, defined by a row of pixels, from the \citet{Protopapa+2017} \N2 map (excluding pixels that fall within SP), and assign every pixel on the non-encounter hemisphere in that row this mean value. The final map combines these as follows: on the encounter hemisphere, we assume the product of the \citet{Lewis+2019} \N2 presence map and the \citet{Protopapa+2017} fractional abundance map, while on the non-encounter hemisphere we assume the product of the Gabasova et al. (2020) \N2 presence map and the \citet{Protopapa+2017} fractional abundance map. Hereafter referred to as the reference map, our assumed \N2 spatial distribution map for latitude north of 35\deg S is shown in Figure \ref{fig:n2}. The fractional abundance of \N2 in each location affects our calculation of the insolation, as described below, and also of the thermal emission, since only the fraction of the area covered by \N2 is assumed to radiate. In Figure \ref{fig:n2}, the grayscale indicates the fractional abundance of \N2 ice, with black indicating \N2-free areas and white indicating 100\% coverage of \N2 ice. In reality, the \N2-covered areas have varying thicknesses of ice, with SP having perhaps 5 km of ice \citep{McKinnon+2016} while the midlatitude deposits may be much thinner. 

We make several different assumptions for the unobserved southern hemisphere (south of 35\deg S). We use (i) a bare southern hemisphere, (ii) a south polar cap extending from the pole to 60\deg S with a fractional abundance of 20\%, and (iii) a southern zonal band of \N2 ice between 35\deg S and 55 \deg S with a fractional abundance of 20\%. We also present results from a simplified case assuming SP contains the only surface deposit of \N2, to emphasize the significant effect of this feature on the global pressure.

\begin{table}[]
	\begin{center}
		\begin{tabular}{|c|c|c|}
			\hline
			& Percentage covered by \N2   & Equivalent Area [m$^2$] \\ \hline
			\hline
			Reference & 10.18\%  & 1.81 x 10$^6$  \\ \hline
			SP-only  & 1.72\% & 1.05 x 10$^5$       \\ \hline
			Southern Polar Cap & 11.52\% & 2.05 x 10$^6$\\ \hline
			Southern Zonal Band & 12.64\% & 2.25 x 10$^6$ \\ \hline
		\end{tabular}
		\caption{Amount of Pluto's surface that is covered by \N2 in each of the four distributions.}
		\label{table:fracs}
	\end{center}
\end{table}

\begin{figure}
	\begin{center}
		\includegraphics[width =0.55 \textwidth]{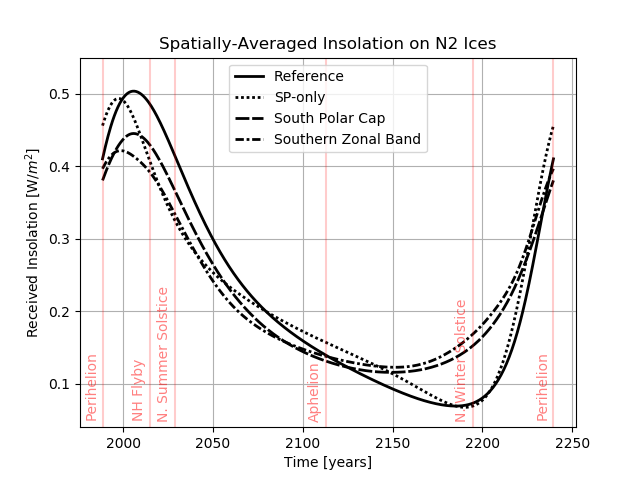}
		\caption{Spatially-averaged insolation onto the \N2 ice in each of our four distributions: reference model (solid line), SP-only model (dotted), south polar cap model (dashed), and southern zonal band (dash-dotted). The x-axis shows time in Earth years spanning one Pluto orbit, beginning in 1988.}
		\label{fig:insol}
	\end{center}
\end{figure}

For each choice of \N2 distribution, we calculate the diurnally- and spatially-averaged insolation onto the surface ices as a function of time, which is an input to VT3D, as shown in Figure \ref{fig:insol}. Table \ref{table:fracs} shows the amount of the surface that is covered by \N2 in each distribution. Equation A.3 in the Appendix shows how we calculate the diurnally-averaged insolation as a function of latitude and time. From this, we find the spatial-average insolation using equation A.4, taking into account both \N2 presence and the fractional abundance of \N2 in each location. In doing so, we assume that the distributions are static in time, and that the surface ices are in vapor pressure equilibrium with the atmosphere, and can thus be described by a single temperature dependent on the average insolation. We investigate the effect seasonal, mobile \N2 would have on the model results in Section 3.5, finding that it is difficult to match observations with the inclusion of mobile \N2. Assuming a static distribution is a simplification, which allows us to investigate multiple distributions at a lower computational cost, but it is also motivated by the fact that many of Pluto's \N2 ice deposits appear to be perennial (persisting for longer than one orbit). SP is a perennial feature: the surface of the ice sheet is estimated to be less than 10 My old  \citep{White+2017} based on the lack of impact craters, but the ice sheet is undergoing convection with an overturning timescale of 0.5 My which cyclically refreshes the surface, allowing the ice sheet to be much older than the crater-derived age. The underlying basin is ancient and likely greater than 4 Gy old \citep{Moore+2016}. Numerical simulations from \citet{Bertrand2016} found that all of the \N2 ice was sequestered into a 3-km deep SP-like basin within 10,000 Earth years, where it stayed for the remainder of the 50,000-year simulation, strengthening the argument for a perennial SP. It is not as obvious if the other \N2 deposits in the reference map are perennial and last for many Pluto years, or only seasonal and disappear (and reappear) due to sublimation (and condensation) on timescales of tens of Earth years. \N2 is observed at lower altitudes in the northern mid-latitudes (e.g., \citet{Howard+2017}) in depression floors that appear flat and smooth.  This suggests a deeper, perennial \N2 deposit, coating and smoothing underlying rough terrain, rather than an seasonal deposit of a few meters or less \citep{Young+2020AAS}. \citet{Bertrand+2019} showed that the mid-latitude \N2 deposits in the northern hemisphere tend to be seasonal, especially those located within depressions. It is unknown whether \N2 exists at mid to high southern latitudes, and, if it does, whether it is perennial or seasonal.  For computational expediency, we investigate only static southern distributions too.  Here, the term ``static'' refers only to the locations of the \N2 ice; \N2 still sublimes from areas of high insolation and condenses onto areas of low insolation, but initially bare locations and initially \N2 ice-covered locations remain so throughout the length of our models. Future work could relax the requirement of a static distribution and time-constant physical parameters, in order to study various feedback effects, such as condensation of \N2 onto winter latitudes \citep{Hansen+2015, Bertrand+2018}; runaway albedo feedback \citep{Earle+2018}; or the impact of haze on the albedo, emissivity, or thermal inertia. 


In addition to the four distributions discussed above, we investigated two others of interest. The first was to exactly mirror the 35-55\deg N \N2 band in the southern hemisphere, rather than use a uniform fractional abundance for the zonal band. With this distribution, the insolation (and therefore the resulting pressure curves) was nearly indistinguishable from our Southern Zonal Band distribution (which uses a uniform 20\% fractional abundance). Since it is unlikely that the southern hemisphere will be exactly identical to the north in that way (due to local topography, existing substrate albedo, etc.), we chose to present the results from the Southern Zonal Band distribution as described above rather than this symmetric version. We also investigated a distribution in which the entire southern hemisphere exactly mirrors the northern hemisphere (as defined in our Reference distribution), except for SP, which was replaced by \N2 ice with a constant fractional abundance of 0.3 (similar to the surrounding \N2; the average fractional abundance of SP is 0.5-0.6). This produces an insolation vs time pattern that is nearly symmetrical around perihelion, and has the peak insolation occurring at perihelion (1990). With this insolation, it is impossible to recreate the doubling/tripling of the atmospheric pressure seen in observations; most (\textit{A}, $\Gamma$, $\epsilon$) cases we investigated had a ratio of predicted 2015 to 1988 pressures of unity, which is outside the 3-σ range constrained by observations.

\subsection{Parameter Space Search}
For each choice of \N2 distribution, we explore three free parameters: the Bond albedo, \textit{A}, the emissivity, $\epsilon$, and the thermal inertia, $\Gamma$, of the surface \N2 ice. We assume for simplicity that each of these parameters is uniform across all of the \N2 ice and constant in time. We perform a grid search of albedo and thermal inertia values, and use the emissivity value that is required to match the New Horizons radio occultation surface pressure of 11.5$\pm$0.7 $\mu$bar in 2015 \citep{Hinson+2017}. To do so, we start with an initial guess at the emissivity, calculate the 2015 surface pressure, and then use a secant method solver to iteratively find the emissivity value which returns a 2015 pressure of 11.5 $\mu$bar. We explore the full range of Bond albedos (between 0 and 1), and thermal inertias between 25 and 2000 tiu ($J m^{-2} K^{-1} s^{-1/2}$). \citet{Lellouch+2013} calculates diurnal thermal inertias based on TNO observations on the order of 10 tiu, much lower than the annual values we derive for most cases (by ``diurnal thermal inertia'', we mean thermal inertia of the material within the diurnal skin depth, while ``annual thermal inertia'' corresponds to the material within the annual or seasonal skin depth). \citet{Spencer+1992} report thermal inertia values for pure \N2 between 530 and 590 tiu, whether the \N2 is in the $\alpha$- or $\beta$-crystalline phase. On Pluto, the \N2 ices are mixed with some CH$_4$ and CO, lay above an H$_2$O ice substrate ($\Gamma$ = 2100 to 2200 tiu, as reported for Triton in \citet{Spencer+1992}), and could be ``fluffy'', fractured, or otherwise distinct from a pure lab sample of ice. Thus, we explore a wide range of thermal inertia values in this model. 
For each \textit{A}, $\Gamma$, $\epsilon$ triplet we calculate a surface pressure versus time curve using 500 timesteps per orbit. To ensure convergence, we initialize the numerical VT3D model using the analytic approximation as our initial guess, and we run the model over 20 orbits before selecting the final orbit as our result. Details of the analytic approximation are given in the appendix.  

Once we have a grid of pressure curves (one for each \textit{A}, $\Gamma$, $\epsilon$ triplet), we apply two constraints to eliminate some regions of this parameter space. The first is to eliminate any cases where the emissivity required to match the 2015 New Horizons pressure is outside of the range 0.3 $< \epsilon < $ 1. An emissivity greater than unity is unphysical, and we impose a lower bound of 0.3. \citet{Stansberry+1996} use Hapke theory to calculate \N2 emissivity as a function of grain size and temperature, and found that the emissivity remains above 0.3 at temperatures between 20 and 60 K for grains larger than 1 cm. \citet{Protopapa+2017} found the spectra from most \N2-rich regions, especially those with a high \N2 fractional abundance, were consistent with cm-size or larger grains.
The second constraint is observational. Pluto's atmospheric pressure, as sensed by ground-based stellar occultations, roughly doubled or tripled between the discovery of its atmosphere in 1988 and the New Horizons flyby in 2015. Occultations of Pluto have not probed all the way to the surface, so it is uncertain whether or not the surface pressure experienced the same two- to three-fold increase. If we assume that the surface pressure increase during this time period was the same as the 1205-km altitude pressure increase, then we find $3.14 > P_{2015}/P_{1988} > 1.82$ at the 3-$\sigma$ level for the surface pressures \citep{Elliot+2003b, Hinson+2017}. We eliminate any (\textit{A},$\Gamma$,$\epsilon$) triplets where the ratio of our modeled 2015 and 1988 surface pressures is outside of this range. 

\section{Results}
\begin{figure*}
	\begin{center}
		\includegraphics[width = \textwidth]{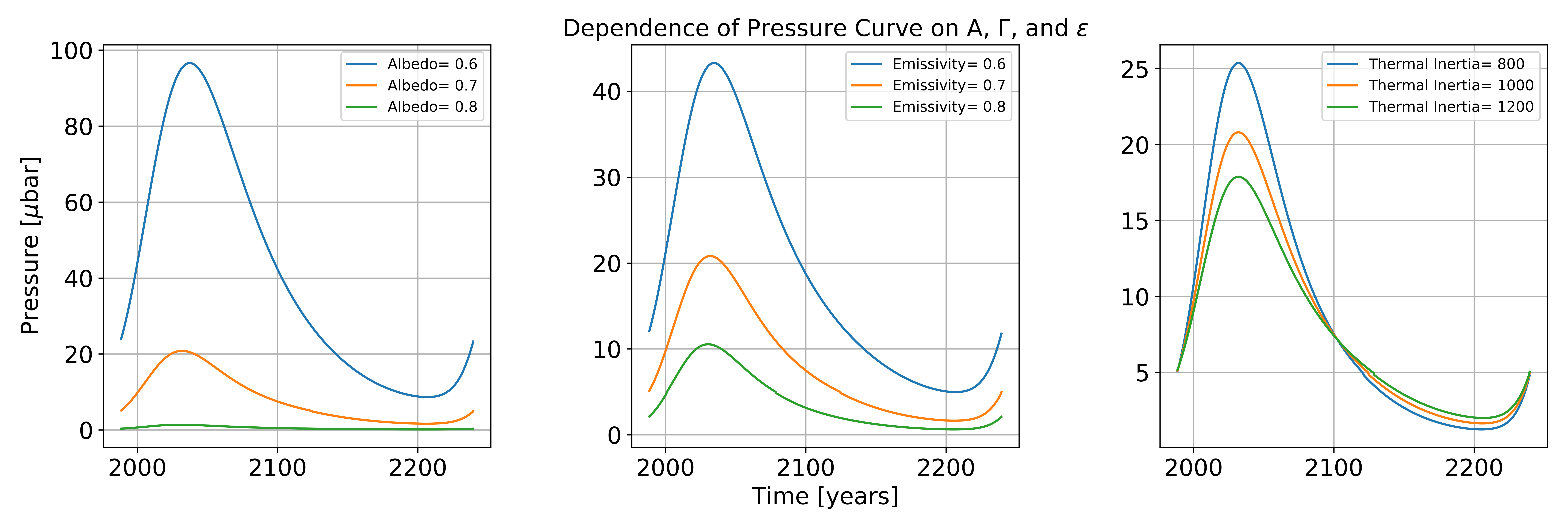}
		\caption{Dependence of the shape of the pressure curve on each of the three free parameters. (Left) Dependence on albedo, for constant thermal inertia of 1000 tiu and emissivity of 0.7. (Center) Dependence on emissivity, for constant albedo of 0.7 and thermal inertia of 1000 tiu. (Right) Dependence on thermal inertia, for constant albedo of 0.7 and emissivity of 0.7.}
		\label{fig:depend}
	\end{center}
\end{figure*}

The dependence of the shape and amplitude of the pressure curve on each of the three parameters is explored in Figure \ref{fig:depend}. The leftmost panel holds the thermal inertia and emissivity constant, at 1000 tiu and 0.7, respectively. For a higher albedo, the resulting pressure is lower at every point in time, due to the lower input of solar energy. The middle panel shows the dependence of pressure on emissivity, while holding albedo constant at 0.7 and thermal inertia at 1000 tiu. The dependence is similar to that of albedo; as emissivity increases the pressure curve is lower at every timestep, as the heat is re-radiated away from the surface more efficiently. The rightmost panel shows how the pressure curve depends on thermal inertia, while albedo and emissivity are both constant at 0.7. A lower thermal inertia surface will experience a larger range of pressures over an orbit compared to a higher thermal inertia one, since the lower thermal inertia surface responds more quickly to changes in the input energy. High thermal inertia materials conduct heat towards the surface more efficiently and thus compensate more efficiently for any change in thermal balance at the surface (e.g. the cooling of the surface by thermal emission). 

In the following sections, we present the annual pressure versus time curves for the wide range of parameter values we explored, for each of our four possible \N2 distributions, and for both Pluto's current orbital configuration and past ``superseasonal'' configurations. We begin with our reference model in Section 3.1, which is the reference map and a bare southern hemisphere. Sections 3.2 through 3.4 present the results from our alternative models, which are (1) a \N2 distribution map where the surface is assumed to be entirely bare except for the \N2 ice contained in SP, (2) the reference map with a south polar cap, (3) the reference map with a southern zonal band. Section 3.5 presents a sensitivity study using mobile \N2 ice. 

\subsection{Reference Model}
We first present the results from Pluto's current orbit using the reference map, along with a bare southern hemisphere. 

\begin{figure}
	\begin{center}
		\includegraphics[width = 0.55 \textwidth]{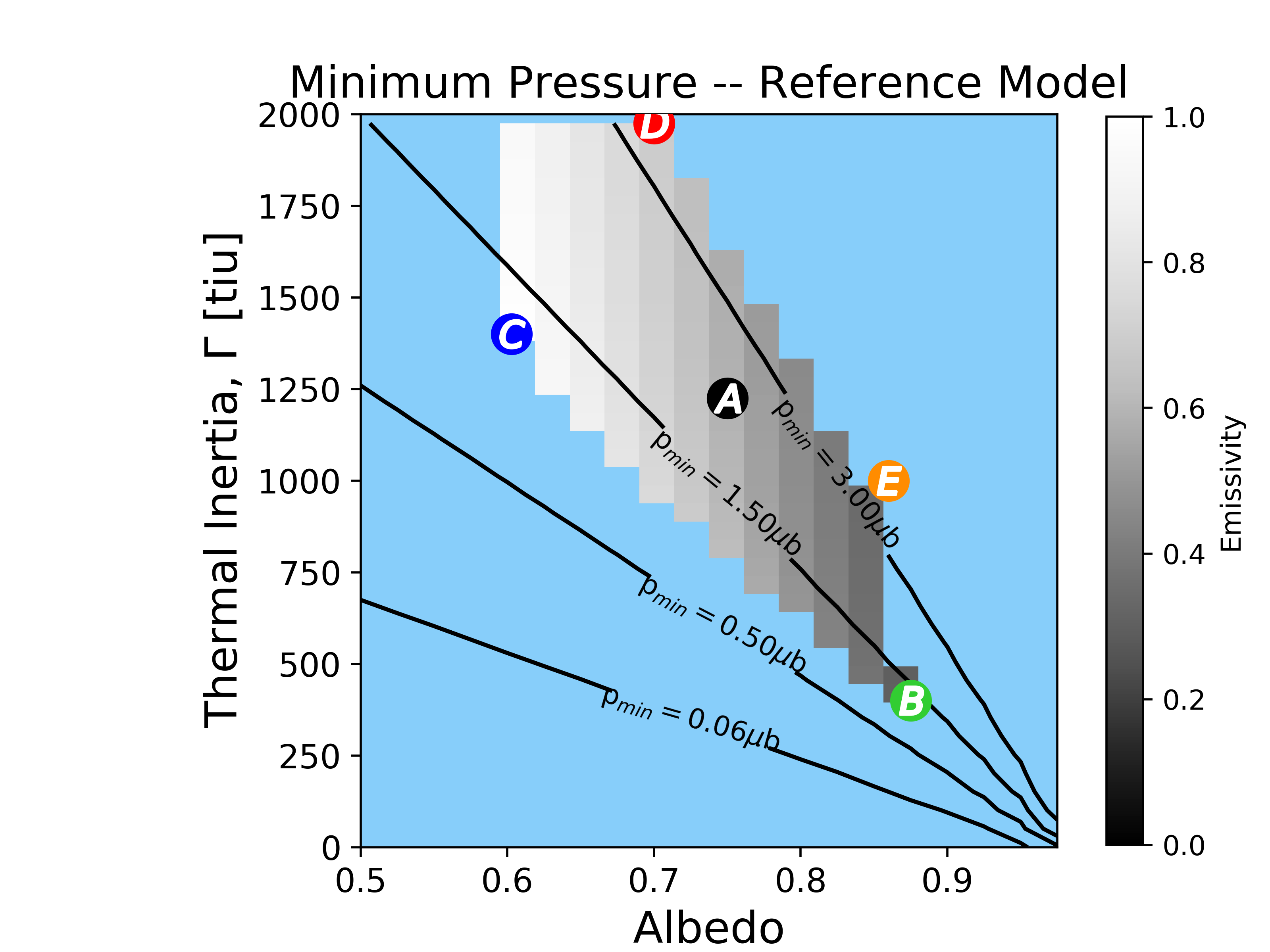}
		\caption{Restricted parameter space for Pluto's current orbit and the reference map (bare southern hemisphere) after choosing $\epsilon$ to ensure $P_{2015} = 11.5$ $\mu$bar, and applying the two further constraints: (1) 1 $> \epsilon >$ 0.3 (2) $3.14 > P_{2015}/P_{1988} > 1.82$. Grayscale indicates the emissivity required and black diagonal contour lines show the minimum pressure experienced over a Pluto year, for that combination of albedo and thermal inertia values. The lettered circles denote the (\textit{A}, $\Gamma$, $\epsilon$) values of the test cases shown in Figures \ref{fig:press}, \ref{fig:extremepress}, and \ref{fig:minpress}, using the same color scheme.}
		\label{fig:grid}
	\end{center}
\end{figure}

\begin{figure}
	\begin{center}
		\includegraphics[width = 0.55 \textwidth]{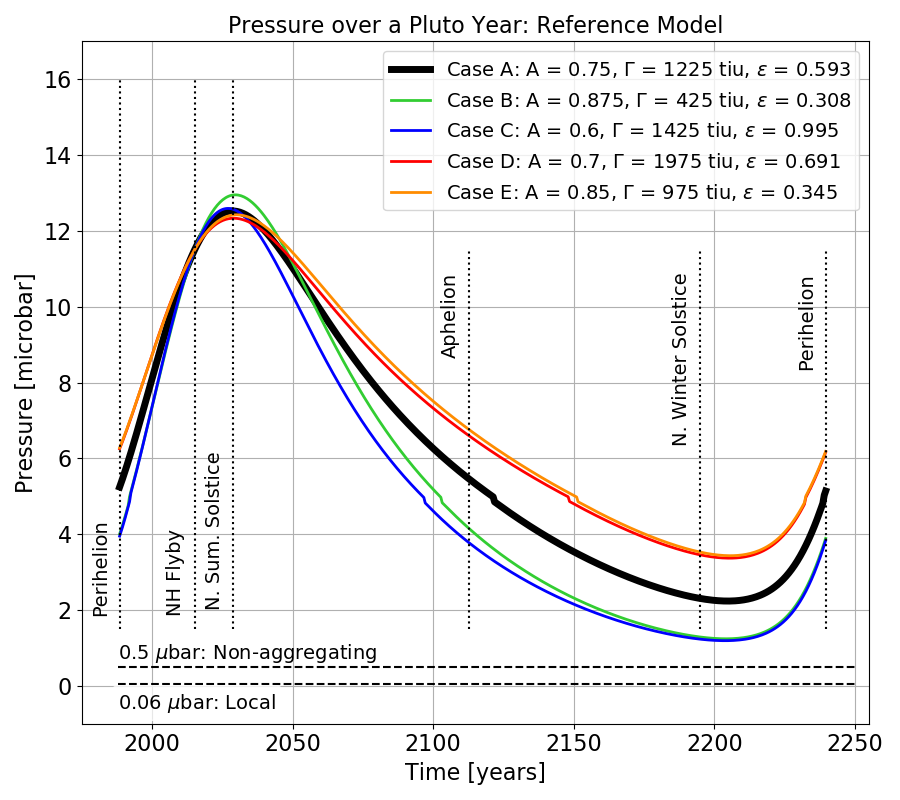}
		\caption{Pressure versus time curves for Pluto's current orbit and the reference map (bare southern hemisphere). The 2\% discontinuity at 5 $\mu$bar reflects the small difference in the calculated pressure at the $\alpha$-$\beta$ transition temperature (see text for details).}
		\label{fig:press}
	\end{center}
\end{figure}

\begin{figure*}
	\begin{center}
		\includegraphics[width = \textwidth]{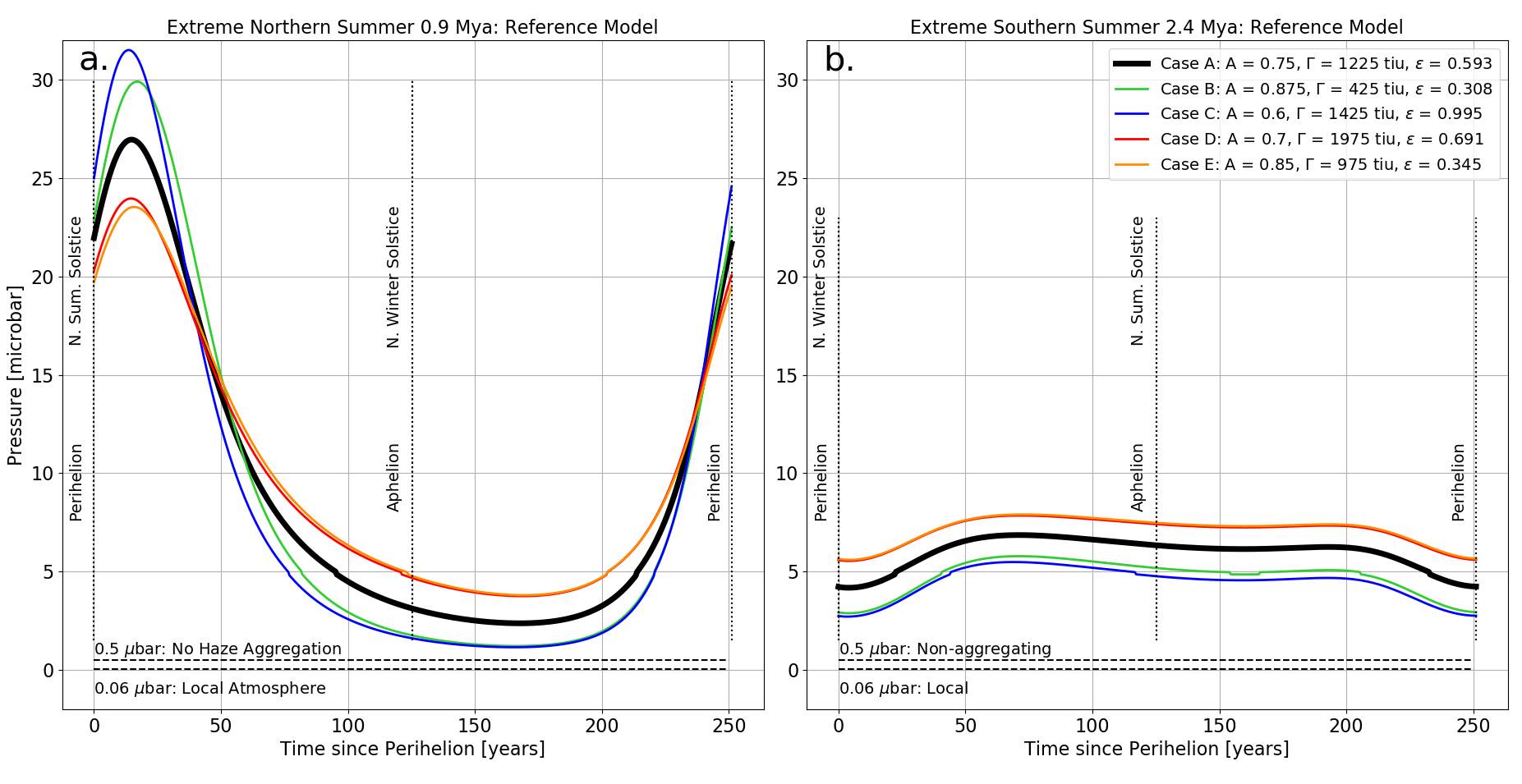}
		\caption{(a) Reference model pressure versus time curves for Pluto's orbit 0.9 Mya, when it was experiencing short, intense northern summers. (b) Reference model pressure versus time curves for Pluto's orbit 2.4 Mya, when it was experiencing long, mild northern summers.}
		\label{fig:extremepress}
	\end{center}
\end{figure*}

\begin{figure}
	\begin{center}
		\includegraphics[width = 0.55 \textwidth]{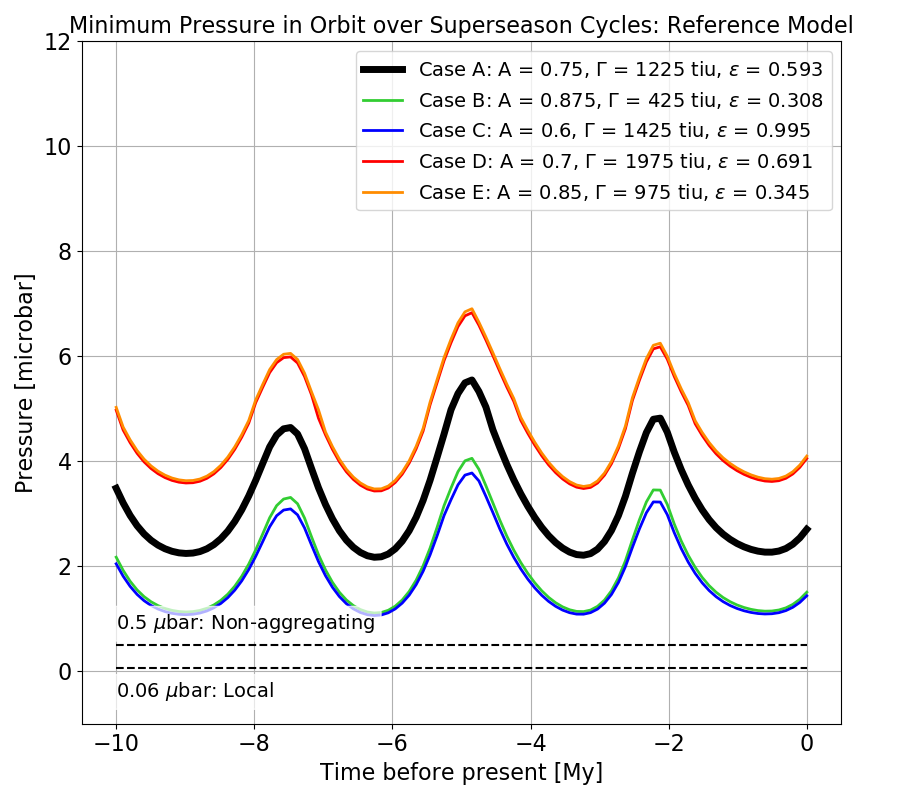}
		\caption{Annual minimum pressure experienced at Pluto's surface over the past 10 My for each of the five test cases, using the reference model.}
		\label{fig:minpress}
	\end{center}
\end{figure}

After applying the constraints as described above for the reference model pressure curves, the remaining allowed parameter space is shown as the grayscale boxes in Figure \ref{fig:grid}. Albedos between 0.6 and 0.9 and thermal inertias above 400 tiu satisfy the constraints, with lower albedos requiring higher thermal inertias. All of the cases that had allowable emissivity values and pressure increases between 1988 and 2015 had minimum pressures between 1 and 3.5 $\mu$bar. There are no (\textit{A},$\Gamma$,$\epsilon$) triplets that drop below the 0.5 $\mu$bar haze aggregation limit or the 0.06 $\mu$bar local atmosphere limit, or the even lower atmospheric transparency limit for Pluto's current orbit.

The pressure curves for five example cases, chosen to span the full range of allowable parameter space, are shown in Figure \ref{fig:press}. The thick black line (case A) shows a central case with A = 0.75, $\Gamma$ = 1225 tiu, and $\epsilon$ = 0.593. Case A shows an increase in pressure between perihelion and the peak of pressure shortly after the time of the New Horizons flyby, and then a slow decrease to the minimum pressure near northern winter solstice. The delay between perihelion and the peak of pressure is primarily due to the subsolar latitude dependence. The \N2 ices receive the strongest spatially-averaged insolation near 2008 (see Figure \ref{fig:insol}), which is determined in part from the 1/r$^2$ dependence but more strongly depends on the incidence angle of sunlight onto SP. Thermal inertia adds to this delay as well. The jump in pressure near 5 $\mu$bar present in all five of the curves is caused by the small numerical discontinuity of 2\% at the change in the form of the vapor pressure equation at the $\alpha$-$\beta$ transition of \N2, which occurs at 35.6 K \citep{FraySchmitt+2009}. 

The blue and green curves (cases B and C) are example cases that remain colder (and therefore have a lower surface pressure) than case A throughout most of the orbit. The combination of case B's higher albedo and low thermal inertia compensate for the effect of the low emissivity, keeping the surface colder than in case A. Case C has a lower albedo and a higher emissivity (so it effectively reradiates away the insolation), causing it to be consistently colder. The red and orange curves (cases D and E) in Figure \ref{fig:press} are example cases that remain warmer than case A throughout most of the orbit. Case D has a similar albedo and emissivity as case A, but experiences a smaller range of pressures due to the higher thermal inertia. Case E has a higher albedo than case A and a lower emissivity, so it is able to remain warmer despite a lower thermal inertia by reradiating the input solar insolation less effectively. None of the test cases predict pressures below any of the haze-important pressures; the atmosphere never becomes non-aggregating, local, nor UV-transparent. This reference model predicts a maximum in the pressure between 2027 and 2030, after which the surface pressure will begin to decrease.

As evident in Figure \ref{fig:press}, extrema in the surface pressure occur close to solstices, when the primarily-northern \N2 deposits are receiving the most (or least, in the case of winter solstice) direct insolation. If northern summer solstice occurs near perihelion, the \N2 deposits will be receiving the most direct insolation (smallest incidence angle) when they are also receiving the most intense insolation (closest to the Sun), creating a strong but short northern summer. Conversely, if northern summer solstice occurs near aphelion, they will be receiving the most direct insolation (smallest incidence angle) when they are receiving the least intense insolation (farthest from the Sun), creating a mild but long northern summer. In order to investigate Pluto's pressure during these extreme seasons, we used the same five example cases as the current orbit and ran VT3D back 10 My, adjusting the obliquity, eccentricity, and subsolar latitude at perihelion according to \citet{Earle+2017}. Figure \ref{fig:extremepress} shows the pressure versus time curve for the five example cases using our reference model during a period of intense northern summer 0.9 Mya (panel a) and a period of intense southern summer (and hence mild northern summer) 2.4 Mya (panel b). The color scheme and labelling of the cases remains the same as Figures \ref{fig:grid} and \ref{fig:press}. 

Figure \ref{fig:extremepress}a shows the extreme summer characteristic of the orbital configuration Pluto was in 0.9 Mya, with a sharp peak just after perihelion and a wide, low minimum in the pressure curve. The pressure varies wildly over an orbit, ranging between 2.5 and 27 $\mu$bar for case A. Despite this wide range, none of the example cases drop below any of the pressures important to haze production, so haze would not be affected during this time period. Note that the minimum pressure does not occur simultaneously with aphelion and northern winter solstice, but rather occurs some time afterward due to the effect of thermal inertia. 

During the mild northern summer at 2.4 Mya shown in Figure \ref{fig:extremepress}b, the pressure curves are noticeably flatter than the 0.9 Mya configuration and have a long peak-plateau where the pressure is stable. Since the reference model assumes a bare southern hemisphere (south of 35\deg S), the southern summer is not particularly extreme; at perihelion/southern summer solstice, the spatially-averaged insolation is very low since no \N2 deposits are receiving direct insolation, which causes the pressure to be low as well. In this configuration, like the current orbit and 0.9 Mya, none of the example cases become cold enough to disrupt haze.

Figure \ref{fig:minpress} shows the minimum pressure experienced over an orbit for the past 10 My (roughly three full obliquity cycles) for the five example cases. None of these curves fall below the 0.5 $\mu$bar nor the 0.06 $\mu$bar levels, or the even lower atmospheric transparency pressure levels. Depending on the choice of albedo, thermal inertia, and emissivity, this model predicts a minimum pressure over the past 10 My between 1 and 4 $\mu$bar. Note that the extreme values in Figure \ref{fig:minpress} do not occur exactly at the superseason epochs 0.9 and 2.4 Mya. This is due to the time offset between winter solstice and the minimum pressure (a seasonal thermal inertia effect).


\subsection{Sputnik Planitia-only Model}
\begin{figure}
	\begin{center}
		\includegraphics[width = \textwidth]{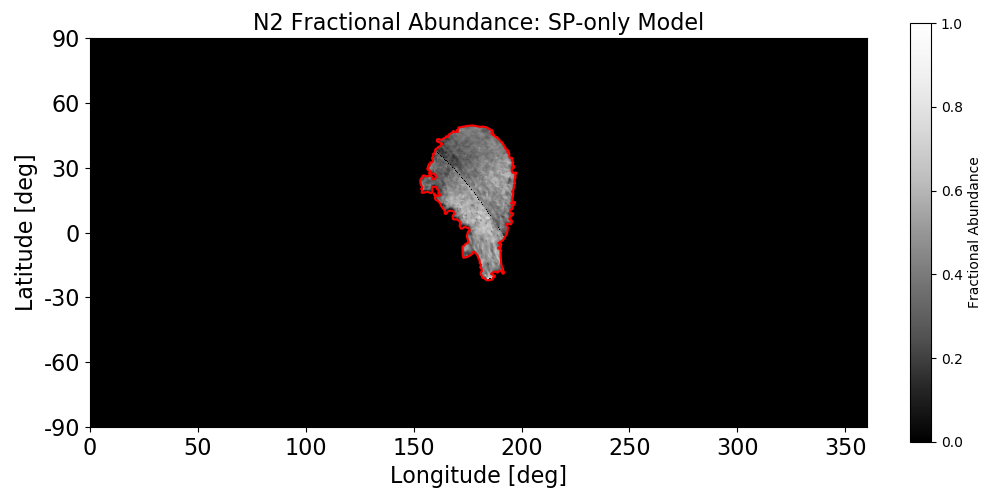}
		\caption{Assumed spatial distribution of \N2 ice for the SP-only model. The red outline shows the boundary of SP as defined by \citet{White+2017}.}
		\label{fig:sp_distribution}
	\end{center}
\end{figure}

Next, we discuss the results from our alternative models, beginning with a \N2 distribution in which SP is the only source of \N2 on the surface. Figure \ref{fig:sp_distribution} shows the \N2 distribution for this alternative model. Both the band depth map \citep{Schmitt+2017} and the Hapke modeling map \citep{Protopapa+2017} clearly indicate deposits of \N2 ice outside of SP, but by limiting this distribution to SP alone, we can investigate the relative influence of SP on the climate compared to the other \N2 deposits. SP is 1000 km in diameter (covering 5\% of Pluto's total surface area), estimated to be 4 to 10 km thick, and has a fractional \N2 abundance as high as 60\%, meaning that as much as 60\% by area of each pixel is covered by \N2 \citep{Protopapa+2017}. SP is located near the equator, spanning from 20\deg S to 50\deg N, so it remains at least partially illuminated for the full range of subsolar latitudes experienced over an orbit. For these reasons, we expect SP to be a strong driver of the atmospheric pressure, and thus expect the SP-only model results to be very similar to the reference model results. This distribution also allows a more direct comparison with \citet{Bertrand2016}, in which \N2 was sequestered into a circular SP-analog basin very similar to this distribution.

\begin{figure}
	\begin{center}
		\includegraphics[width =0.55 \textwidth]{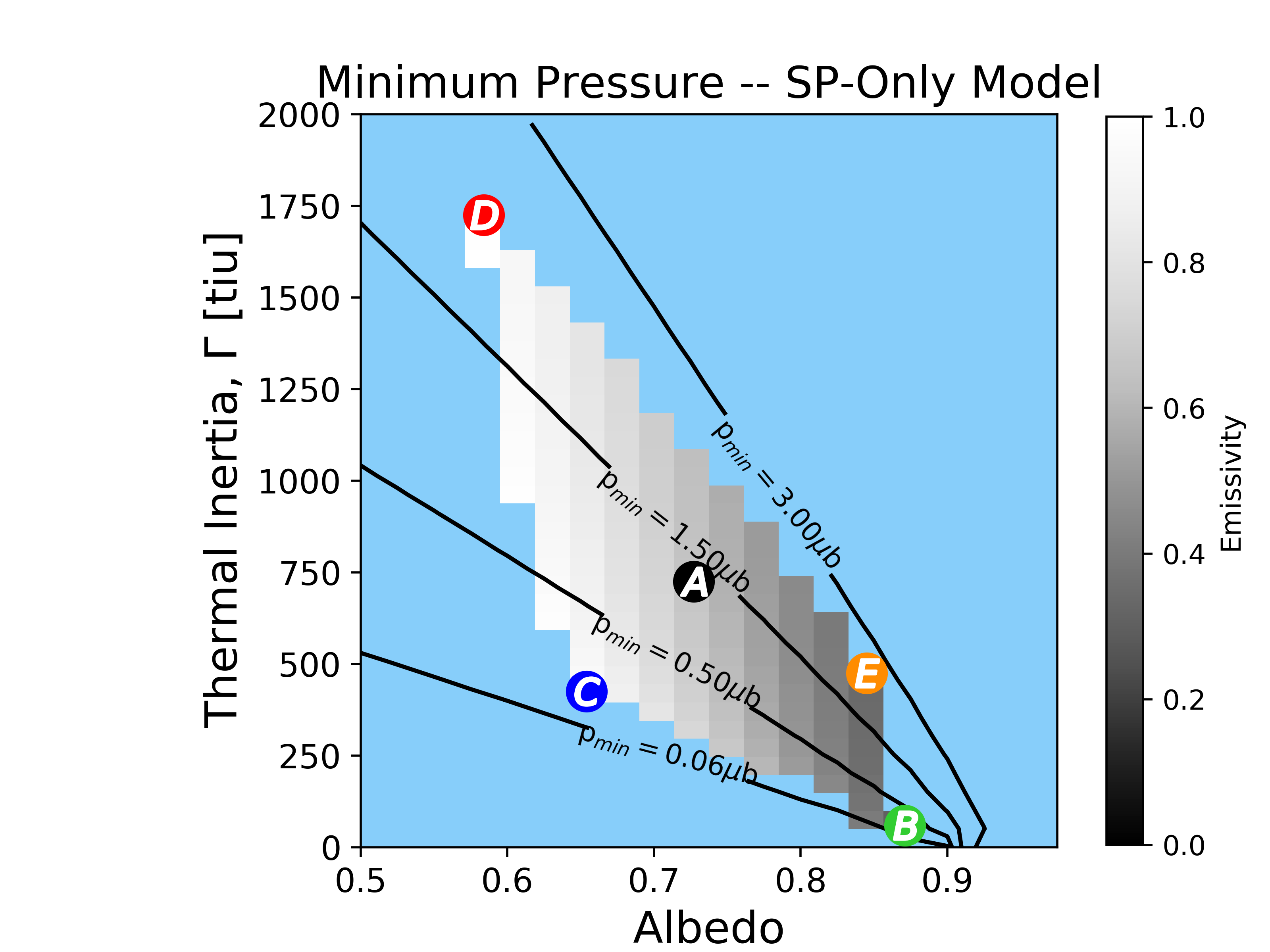}
		\caption{Restricted parameter space for Pluto's current orbit assuming SP is the only \N2 ice deposit, after choosing $\epsilon$ to ensure $P_{2015} = 11.5$ $\mu$bar, and applying the two further constraints: (1) 1 $> \epsilon >$ 0.3 (2) $3.14 > P_{2015}/P_{1988} > 1.82$. Grayscale indicates the emissivity required and black diagonal contour lines show the minimum pressure experienced over a Pluto year, for that combination of albedo and thermal inertia values. The lettered circles denote the (\textit{A}, $\Gamma$, $\epsilon$) values of the test cases shown in Figures \ref{fig:press_sp} and \ref{fig:minpress_sp}, using the same color scheme.}
		\label{fig:grid_sp}
	\end{center}
\end{figure}

\begin{figure}
	\begin{center}
		\includegraphics[width =0.55 \textwidth]{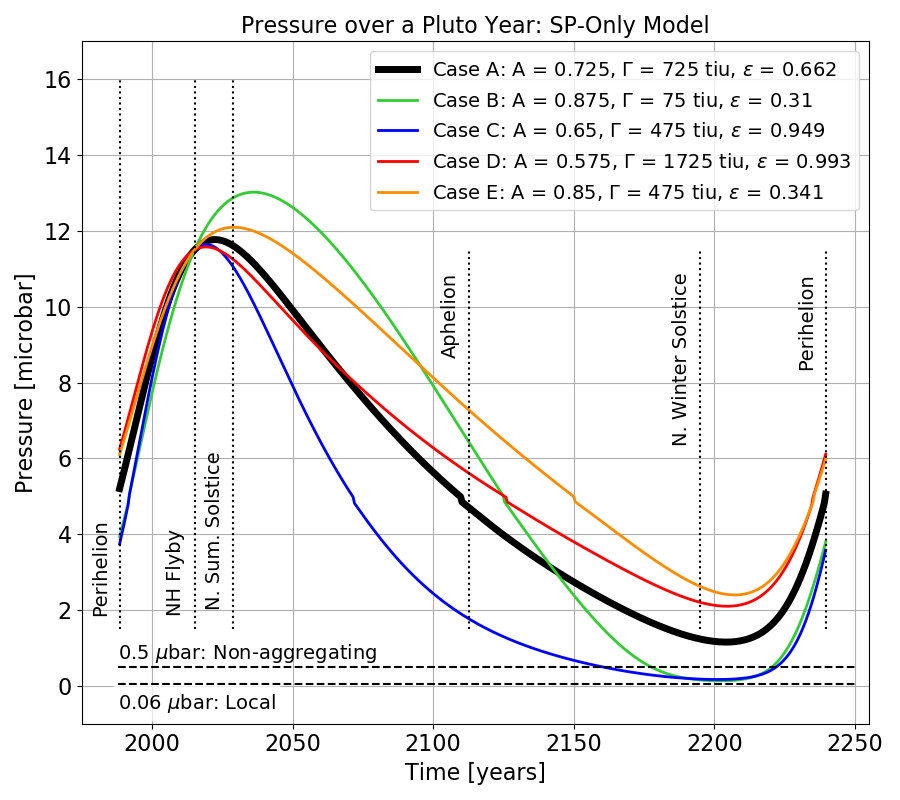}
		\caption{Pressure versus time curves for Pluto's current orbit, assuming SP is the only \N2 ice deposit.}
		\label{fig:press_sp}
	\end{center}
\end{figure}

\begin{figure}
	\begin{center}
		\includegraphics[width =0.55 \textwidth]{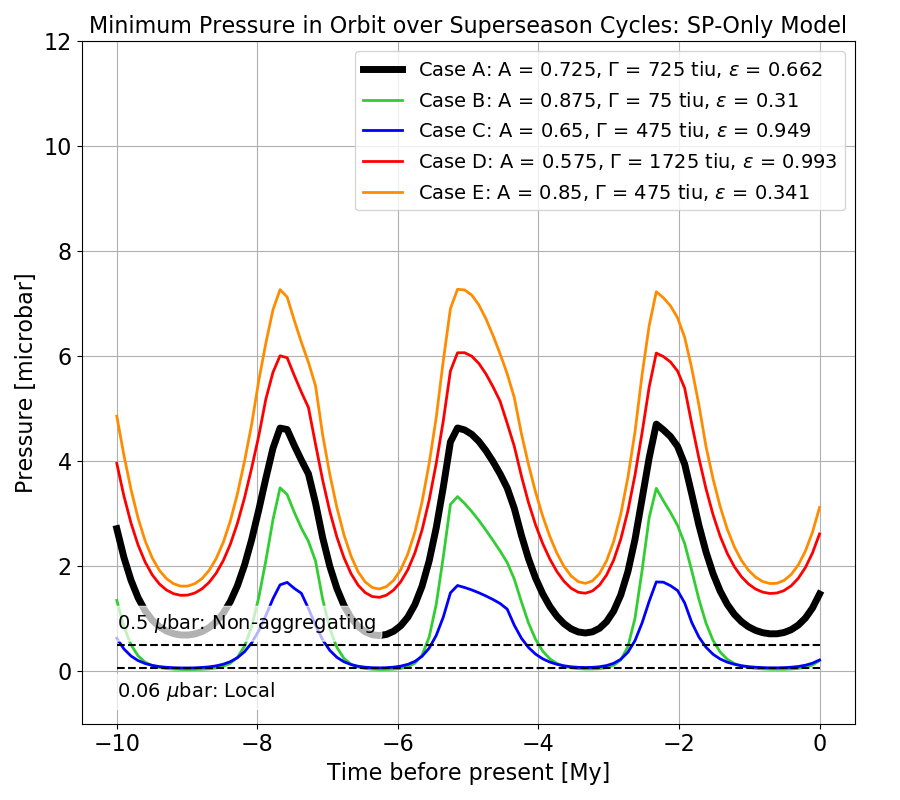}
		\caption{Annual minimum pressure experienced at Pluto's surface over the past 10 My for each of the five test cases, assuming SP is the only \N2 ice deposit.}
		\label{fig:minpress_sp}
	\end{center}
\end{figure}

Figure \ref{fig:grid_sp} shows the restricted parameter space for the SP-only model. In comparison with Figure \ref{fig:grid} for the reference model, lower thermal inertias are required for the SP-only model. Ignoring all of the \N2 ice outside of SP causes the peak in the spatially-averaged insolation to occur sooner after perihelion, and for the difference between the peak value and the perihelion value of the spatially-averaged insolation to be smaller (see Figure \ref{fig:insol}). As a consequence of these two changes to the insolation, lower thermal inertias are needed to compensate, in order to satisfy the constraint on the modeled increase in pressure between 1988 and 2015.

Five example test cases are shown in Figure \ref{fig:press_sp} for the SP-only case. Note that due to the different constrained parameter space, these 5 cases are different than the test cases from the reference model, but the color scheme is the same, with red and orange curves being relatively warmer and higher pressure cases, while the blue and green curves are cooler and therefore lower pressure for much of the orbit. In the SP-only model, the peaks in most of the test case pressures occur slightly earlier, before northern summer solstice, and are slightly lower at 11.5 $\mu$bar (excluding case B) compared to 12.5 $\mu$bar for the reference model test cases. This is again a consequence of the differences in the spatially-averaged insolation between the reference model and the SP-only model. Additionally, the minima in the pressure curves are relatively lower than the reference model case, with the cases B and C dropping below the haze aggregation limit for a period of time near northern winter solstice. This behavior is a consequence of the lower thermal inertias required for the SP-only case: lower thermal inertia allows input energy variations to be quickly realized as temperature variations, creating larger temperature and pressure swings. As expected, the general pressure evolution trend is very similar for the SP-only model compared to the reference model, confirming our expectation that SP is a large driver of the seasonal pressure cycle on Pluto. None of the test cases predict a local atmosphere in Pluto's current orbit.

We investigated the long-timescale behavior of the SP-only model as well. Figure \ref{fig:minpress_sp} shows the minimum pressure experienced in each orbit going back 10 My, for the same five test cases. Test case B produced some past atmospheres that could have been non-aggregating and local, predicting minimum pressures that fall just below 0.06 $\mu$bar for select orbits over the past 10 My. None of the other test cases ever predict non-aggregating or local atmospheres, meaning the modeled atmospheres with those thermal parameters never collapse over the past 10 My (although case C comes very close with minimum pressure of 0.062 $\mu$bar).

\subsection{South Polar Cap Model}
Existing models have shown that perennial polar caps are not likely to form on Pluto, due to the high obliquity which causes the poles to receive more annually-averaged insolation than the equator \citep{Young2013, Bertrand+2018, Bertrand+2019}. Prior to the flyby, \citet{Young2013} found that perennial northern volatiles were possible, but that most perennial southern volatile cases could be eliminated based on the modeled pressure increase between 1988 and 2006 not matching the observed increase from occultations. While the simulations of \citet{Bertrand+2019} did not produce perennial polar caps of \N2, many of their simulations (representing a range of thermal inertia and albedo values for the \N2 ice, CH$_4$ ice, and H$_2$O substrate) resulted in the formation of a seasonal south polar cap that persisted for 80\% to 90\% of Pluto's orbit. Observations by New Horizons found the north polar region north of 60\deg N to be relatively \N2-free, with band depths less than 0.005 and fractional abundances less than 30\% \citep{Schmitt+2017, Protopapa+2017}. The south polar region was experiencing polar night and was thus unobservable. 

\citet{HansenPaige1996} found that southern polar caps persist for a greater fraction of the orbit than northern caps, due to the fact that northern summer occurs as Pluto is approaching perihelion (causing rapid sublimation of the north polar cap and subsequent rapid condensation on to the cold southern polar cap), while southern summer occurs when Pluto is approaching aphelion (causing slower sublimation of the southern polar cap and slower condensation onto the northern polar cap). Their model assumed a small \N2 inventory (50 kg/m$^2$), as did \citet{Young2013}, while the global equivalent layer implied by the presence of SP alone (5 km deep, 1000 km in diameter) is on the order of 10$^5$ kg/m$^2$. A larger \N2 inventory could mean that polar caps grow thick enough to avoid completely sublimating away during the summer, producing perennial polar caps.
 
Normal reflectance maps produced from Pluto-Charon mutual events in the late 1980s showed a bright south polar cap \citep{YoungBinzel1993}. This cap was not necessarily composed of \N2 ice (it could have been bright CH$_4$ ice as well), but it is evidence that at least seasonal southern caps form on Pluto. Additionally, \citet{GrundyFink1996} analyzed 15 years of visible-wavelength spectroscopy (1980-1994) and found that the spectra were consistent with a model in which much of the southern hemisphere (from the pole to 50\deg S) is covered with a \N2-dominated mix of ices, although other solutions could not be conclusively ruled out.

\begin{figure}
	\begin{center}
		\includegraphics[width = \textwidth]{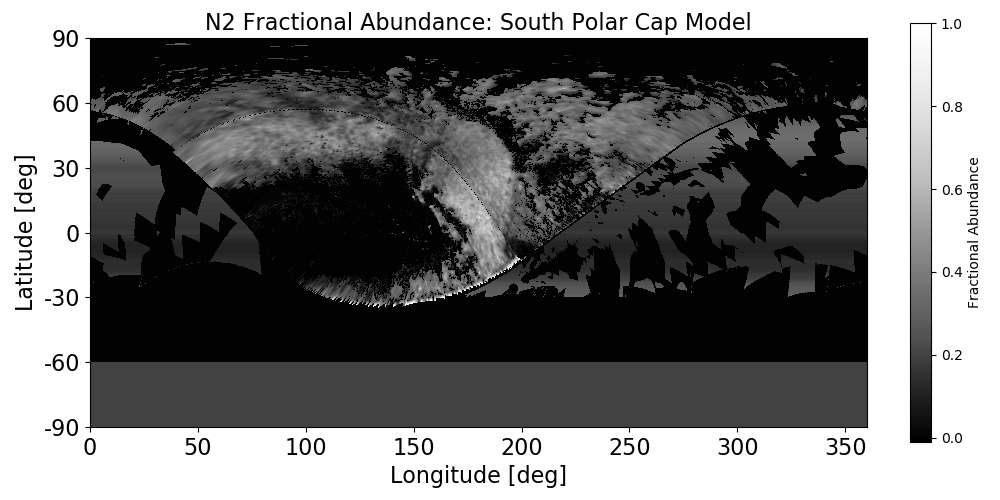}
		\caption{Spatial distribution of \N2 ice for the south polar cap model. Assumes a south polar cap is present extending from the pole to 60\deg S with a fractional abundance of 20\%, in addition to the \N2 present in the reference map.}
		\label{fig:cap}
	\end{center}
\end{figure}

From the above evidence, we do not rule out the possibility of a perennial south polar cap, or a very long-lasting seasonal south polar cap, and choose to investigate it as one of our alternative models. For our south polar cap, we assume a cap of \N2 ice that extends from the pole to 60\deg S with a uniform fractional abundance of 20\%, as shown in Figure \ref{fig:cap}. We investigated polar caps with higher fractional abundances, but found that for larger southern deposits of \N2 ice there were no (\textit{A},$\Gamma$,$\epsilon$) capable of satisfying our constraints.

\begin{figure}
	\begin{center}
		\includegraphics[width =0.55 \textwidth]{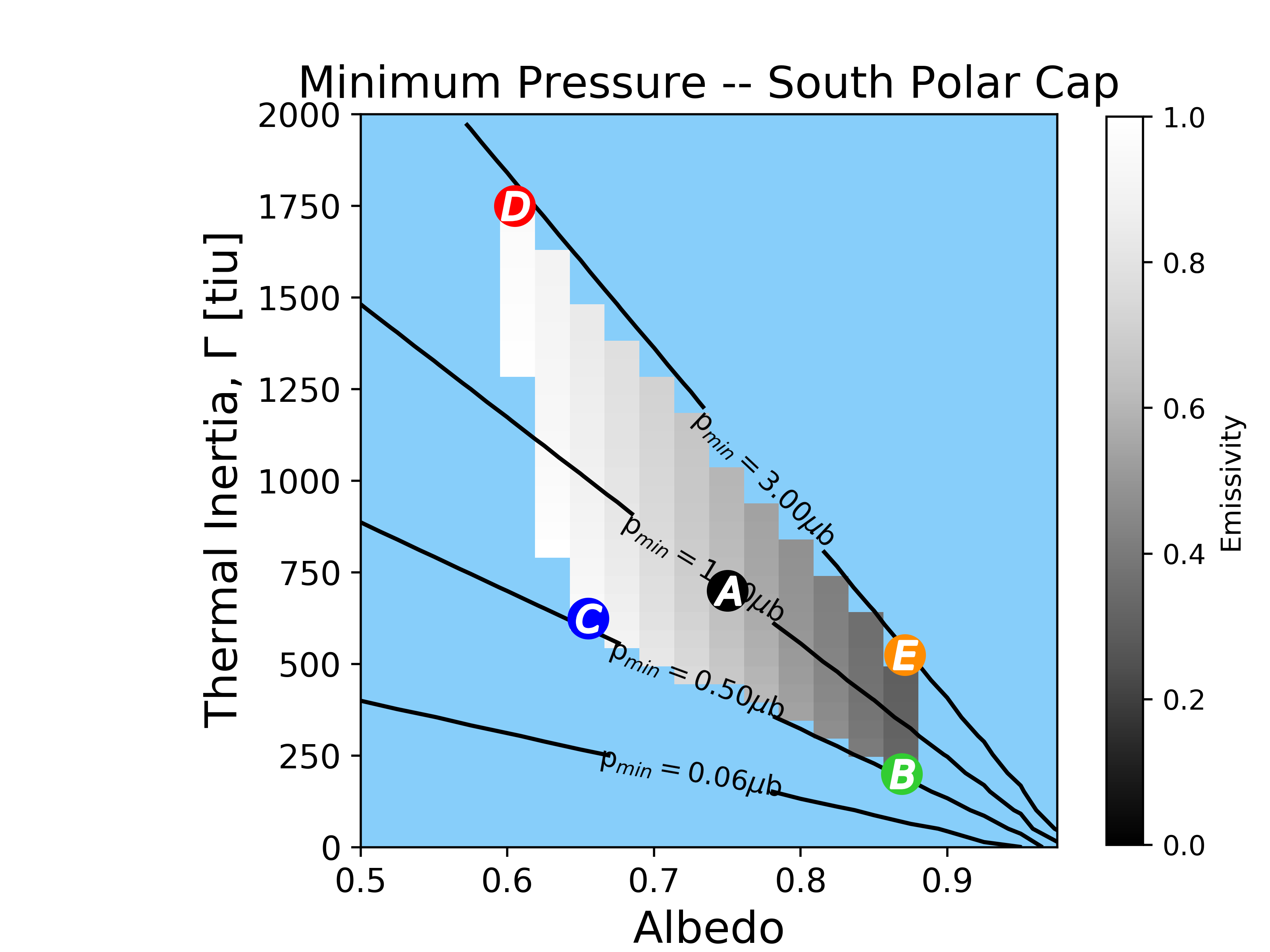}
		\caption{Restricted parameter space for Pluto's current orbit and a south polar cap after choosing $\epsilon$ to ensure $P_{2015} = 11.5$ $\mu$bar, and applying the two further constraints: (1) 1 $> \epsilon >$ 0.3 (2) $3.14 > P_{2015}/P_{1988} > 1.82$. Grayscale indicates the emissivity required and black diagonal contour lines show the minimum pressure experienced over a Pluto year, for that combination of albedo and thermal inertia values. The lettered circles denote the (\textit{A}, $\Gamma$, $\epsilon$) values of the test cases shown in Figures \ref{fig:press_cap} and \ref{fig:minpress_cap}, using the same color scheme.}
		\label{fig:grid_cap}
	\end{center}
\end{figure}

The region of allowed parameter space for the south polar cap model is shown in Figure \ref{fig:grid_cap}. Compared the reference model, lower thermal inertias are required. Minimum pressures between 3 $\mu$bar and 0.5 $\mu$bar are predicted. There are no cases which predict pressures below any of the haze-disruption pressures; aggregation is not interrupted (although case C comes close to a non-aggregating atmosphere with a minimum pressure of 0.506 $\mu$bar), the atmosphere remains global, opaque to UV radiation, and does not collapse.

\begin{figure}
	\begin{center}
		\includegraphics[width =0.55 \textwidth]{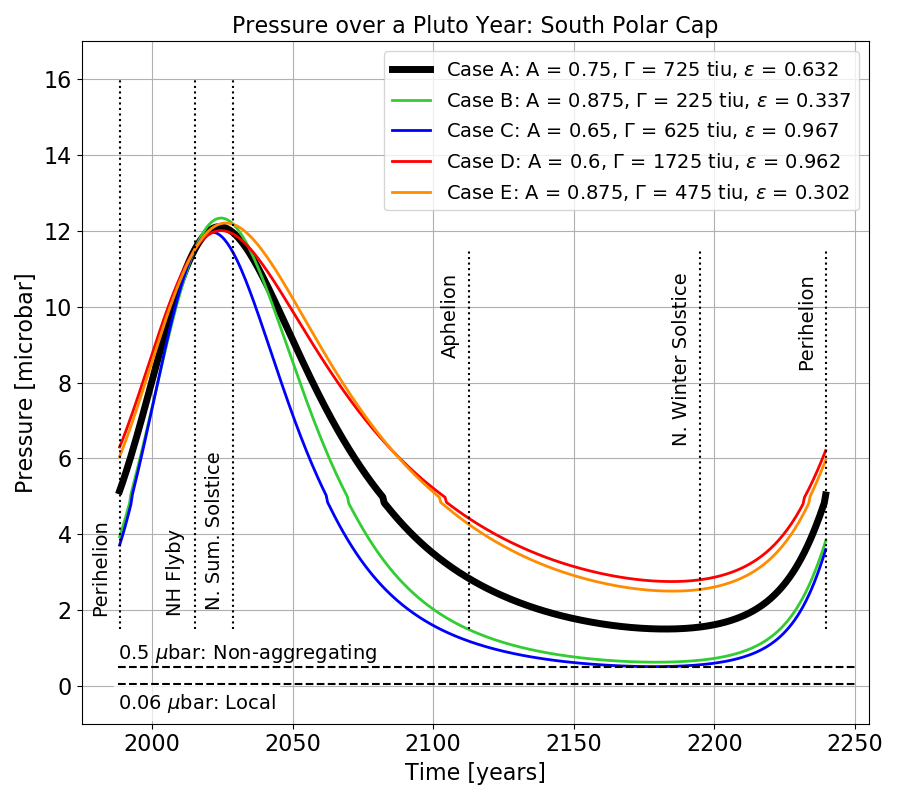}
		\caption{Pressure versus time curves for Pluto's current orbit, for the south polar cap model.}
		\label{fig:press_cap}
	\end{center}
\end{figure}

Five test cases from the region of allowed parameter space are shown in Figure \ref{fig:press_cap}. Overall, the shape and amplitude of the pressure curves are very similar to those from the reference model, with slightly lower maximum and minimum pressures for the south polar cap model. The pressure falls off more quickly in the south polar cap model, leading to a broader minimum extending from aphelion to winter solstice. This behavior, along with the slightly lower maximum and minimum pressures, occurs because the ice in the south polar cap is radiating away energy via thermal emission during the entire orbit (as are the northern hemisphere ices), but is obscured in polar night thus isn't absorbing any solar insolation for part of the orbit. 

\begin{figure}
	\begin{center}
		\includegraphics[width =0.55 \textwidth]{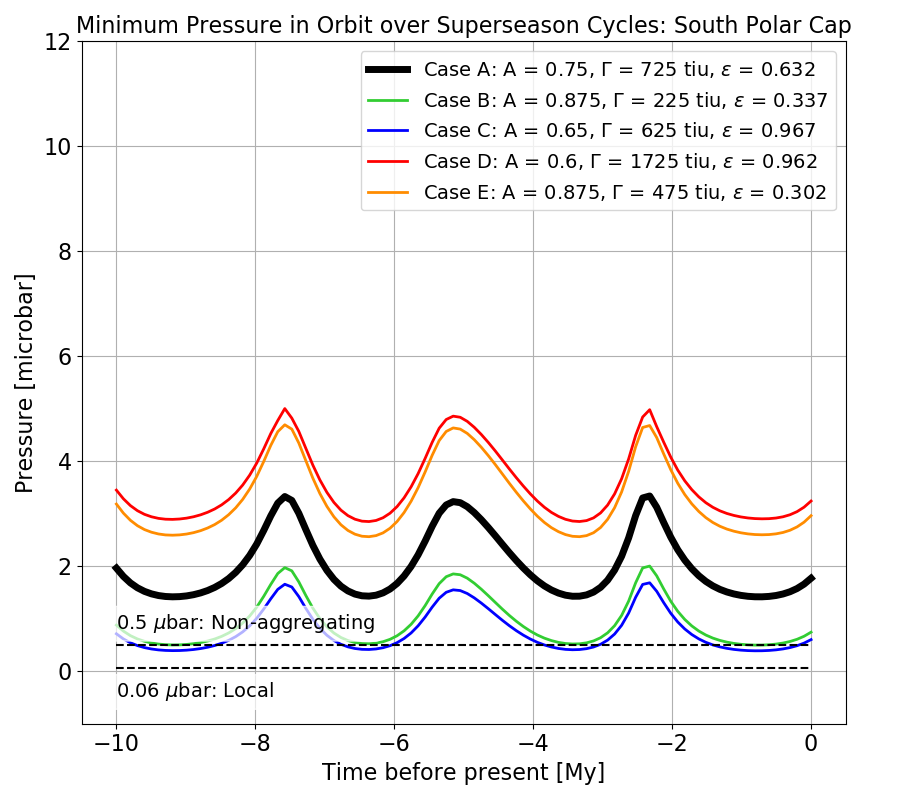}
		\caption{Annual minimum pressure experienced at Pluto's surface over the past 10 My for each of the five test cases, for the south polar cap model.}
		\label{fig:minpress_cap}
	\end{center}
\end{figure}

The superseasonal behavior of the five test cases for the south polar cap model is shown in Figure \ref{fig:minpress_cap}. Cases B (green) and C predict pressures that fall below the haze aggregation limit at points in the obliquity cycle despite remaining above the limit in Pluto's current orbital configuration. Near the extreme northern summer period at 0.9 Mya, the minimum pressure over an orbit predicted in Case C drops to 0.41 $\mu$bar and Case B drops to just under the limit at 0.497 $\mu$bar. In this orbital configuration, the south pole is pointed most directly at the sun at aphelion. The majority of the \N2 ice deposits are not directly illuminated since they are in the northern hemisphere, and despite direct insolation, the \N2 ice at the south pole is not receiving intense insolation due to the high heliocentric distance. Case C has a high emissivity of 0.967, so the unilluminated northern volatiles efficiently reradiate what little solar energy the southern volatiles absorb, causing the low minimum pressure. Case B's high albedo prevents the small amount of northern volatiles from absorbing much energy neear winter solstice, contributing to the low minimum pressure. The other three test cases' combination of albedo, thermal inertia, and emissivity values are able to counteract the orbital configuration's effect on the pressure and their predicted pressures remain above all of the haze-disruption pressures.

\subsection{Southern Zonal Band Model}
\begin{figure}
	\begin{center}
		\includegraphics[width = \textwidth]{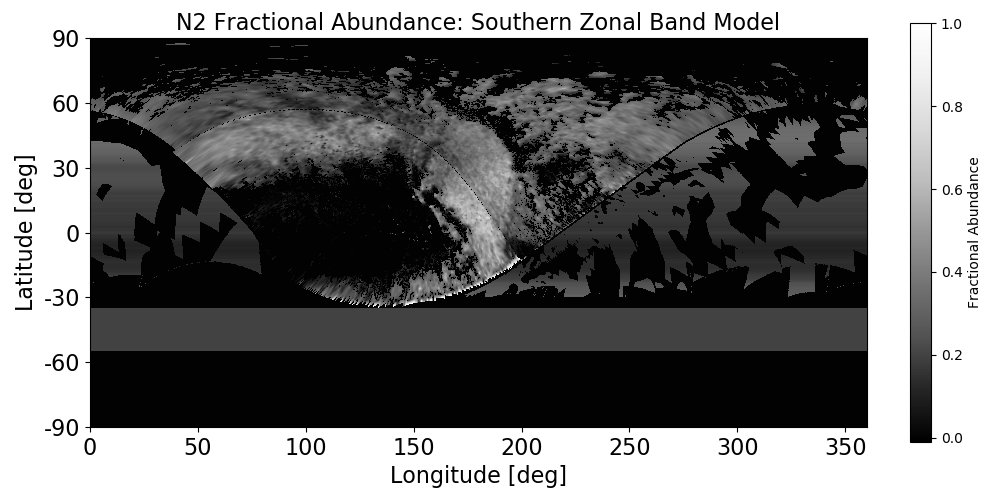}
		\caption{Spatial distribution of \N2 ice for the southern zonal band model. Assumes a zonal band of \N2 is present between 35\deg S and 55\deg S with a fractional abundance of 20\%, in addition to the \N2 present in the reference map.}
		\label{fig:band}
	\end{center}
\end{figure}

Figure \ref{fig:band} shows the \N2 distribution for the final alternative model we investigate, the southern zonal band model. This distribution consists of the reference map plus a zonal band of \N2 between 35\deg S and 55\deg S with a fractional abundance of 20\%. This location was chosen to be similar to the northern midlatitude distribution; between 35\deg N and 55\deg N there is a band of \N2 with an average fractional abundance of roughly 40\%, visible in the reference map and also identified in \citet{Protopapa+2017}. We initially tried a southern zonal band with a fractional abundance of 40\% to match the observed northern band, but found there were no (\textit{A},$\Gamma$,$\epsilon$) triplets capable of satisfying the constraints we imposed. Having such a significant deposit of \N2 ice in the southern hemisphere produced very high spatially-averaged insolation and therefore high pressures in 1988 (near perihelion and equinox). Even with very low thermal inertias ($<$50 tiu), it was not possible to double or triple the atmospheric pressure between 1988 and 2015 while requiring that the modeled 2015 pressure be 11.5 $\mu$bar. This is consistent with results from \citet{Meza+2019}, who found that small southern \N2 deposits (or no southern \N2 at all) were required to produce reasonable pressure evolution in which the peak of pressure occurs after 2015. Thus, we adopt a fractional abundance of 20\% for the southern zonal band. A northern boundary for this band of 35\deg S places it just out of view of the high resolution encounter hemisphere images. At the time of the New Horizons flyby in 2015, everything south of 40\deg S was experiencing polar night.

\begin{figure}
	\begin{center}
		\includegraphics[width =0.55 \textwidth]{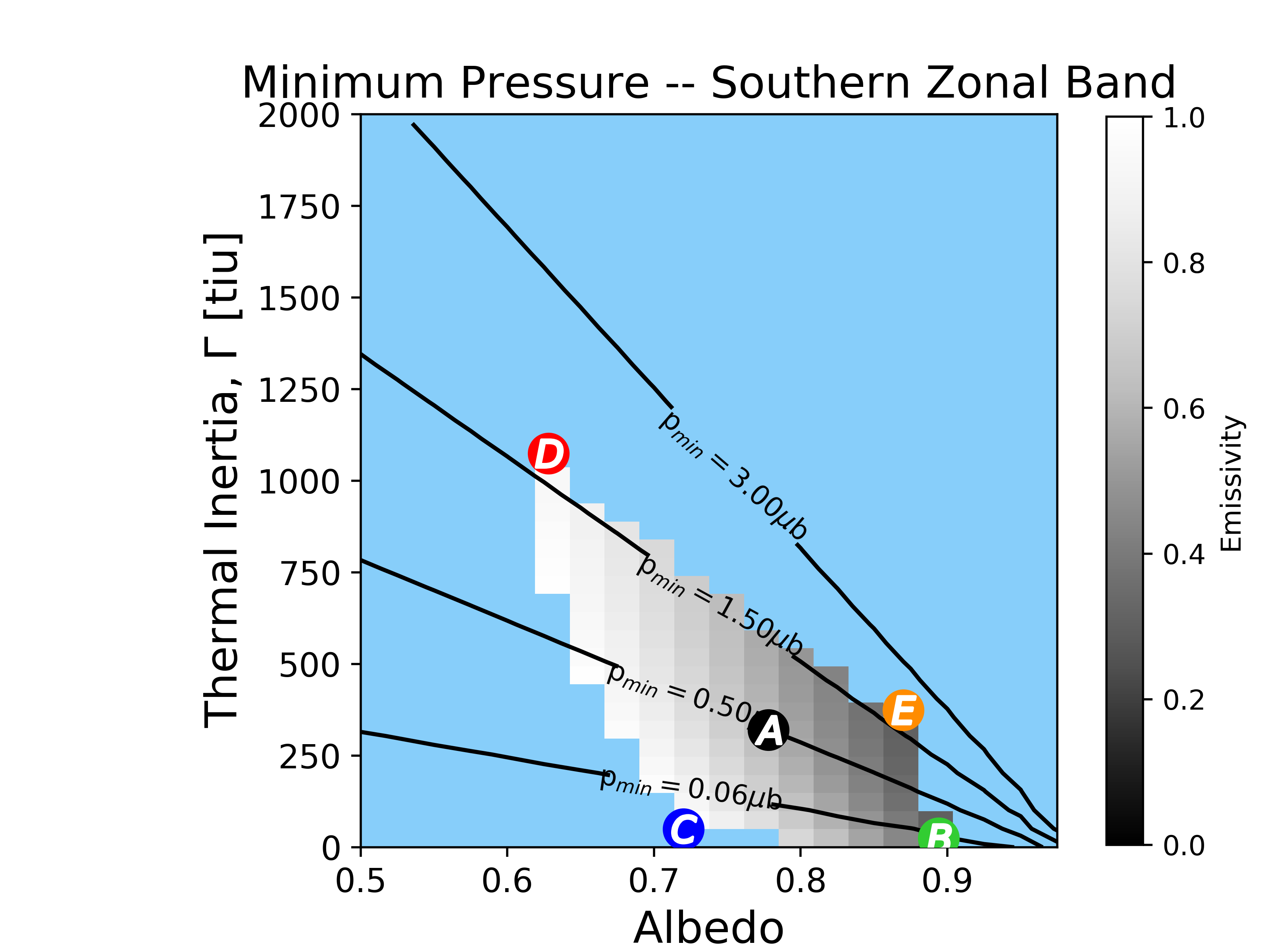}
		\caption{Restricted parameter space for Pluto's current orbit and a southern zonal band after choosing $\epsilon$ to ensure $P_{2015} = 11.5$ $\mu$bar, and applying the two further constraints: (1) 1 $> \epsilon >$ 0.3 (2) $3.14 > P_{2015}/P_{1988} > 1.82$. Grayscale indicates the emissivity required and black diagonal contour lines show the minimum pressure experienced over a Pluto year, for that combination of albedo and thermal inertia values. The lettered circles denote the (\textit{A}, $\Gamma$, $\epsilon$) values of the test cases shown in Figures \ref{fig:press_band}, \ref{fig:press_band_log}, and \ref{fig:minpress_band}, using the same color scheme.}
		\label{fig:grid_band}
	\end{center}
\end{figure}

Figure \ref{fig:grid_band} shows the region of allowed parameter space for the southern zonal band model. Thermal inertias between 25 and 1000 tiu are able to satisfy our constraints. Minimum pressures range between 1.5 $\mu$bar to 0.01 $\mu$bar. Many of the (\textit{A},$\Gamma$,$\epsilon$) triplets produce pressure curves that fall below the haze aggregation limit. Albedos between 0.7 and 0.9 coupled with thermal inertias lower than 200 tiu and nearly the full range of emissivities (0.3 $<$ $\epsilon$ $<$ 1) lead to atmospheric collapse.

\begin{figure}
	\begin{center}
		\includegraphics[width =0.55 \textwidth]{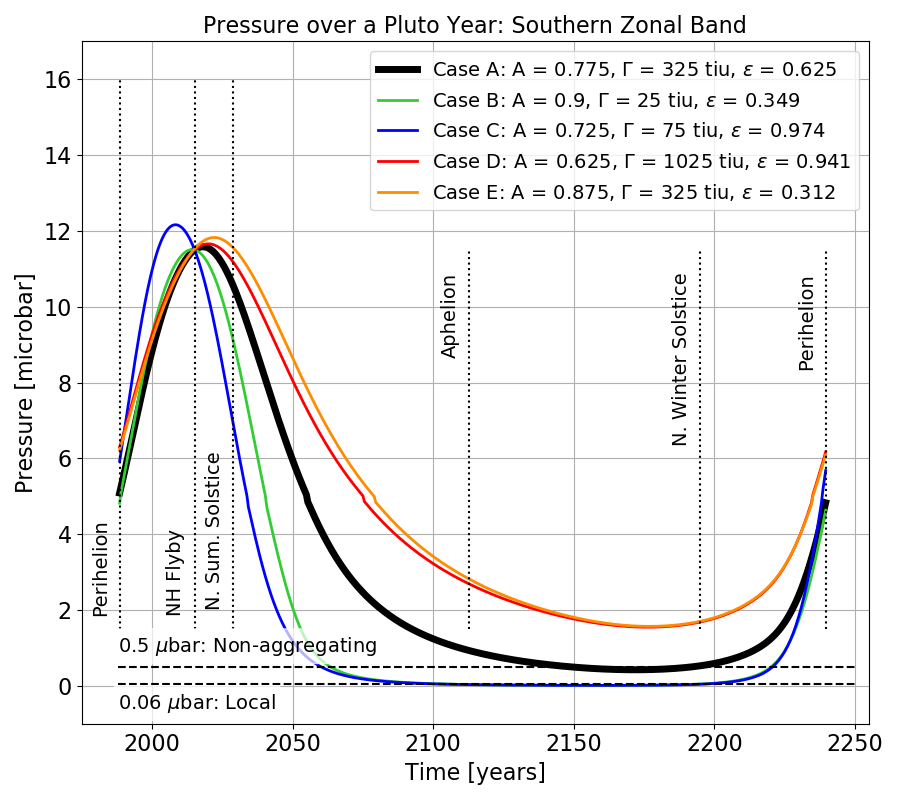}
		\caption{Pressure versus time curves for Pluto's current orbit, for the southern zonal band model (linear scale).}
		\label{fig:press_band}
	\end{center}
\end{figure}

\begin{figure}
	\begin{center}
		\includegraphics[width =0.55 \textwidth]{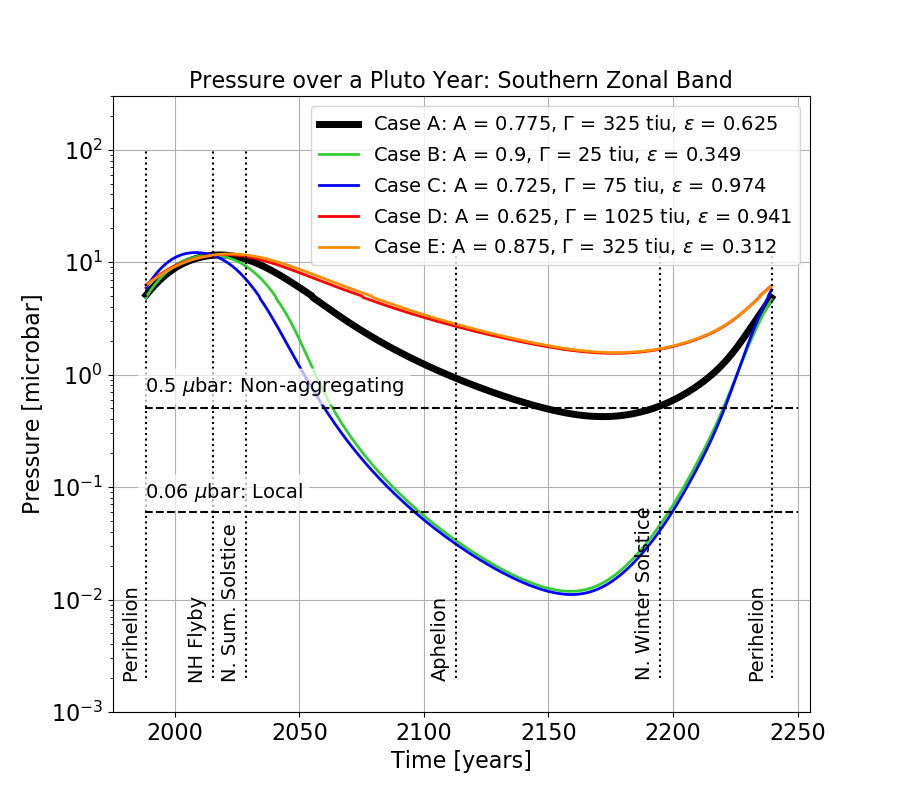}
		\caption{Pressure versus time curves for Pluto's current orbit, for the southern zonal band model (log scale). At this scale, Case B is nearly coincident with Case C, and Case D is nearly coincident with Case E.}
		\label{fig:press_band_log}
	\end{center}
\end{figure}

\begin{figure}
	\begin{center}
		\includegraphics[width =0.55 \textwidth]{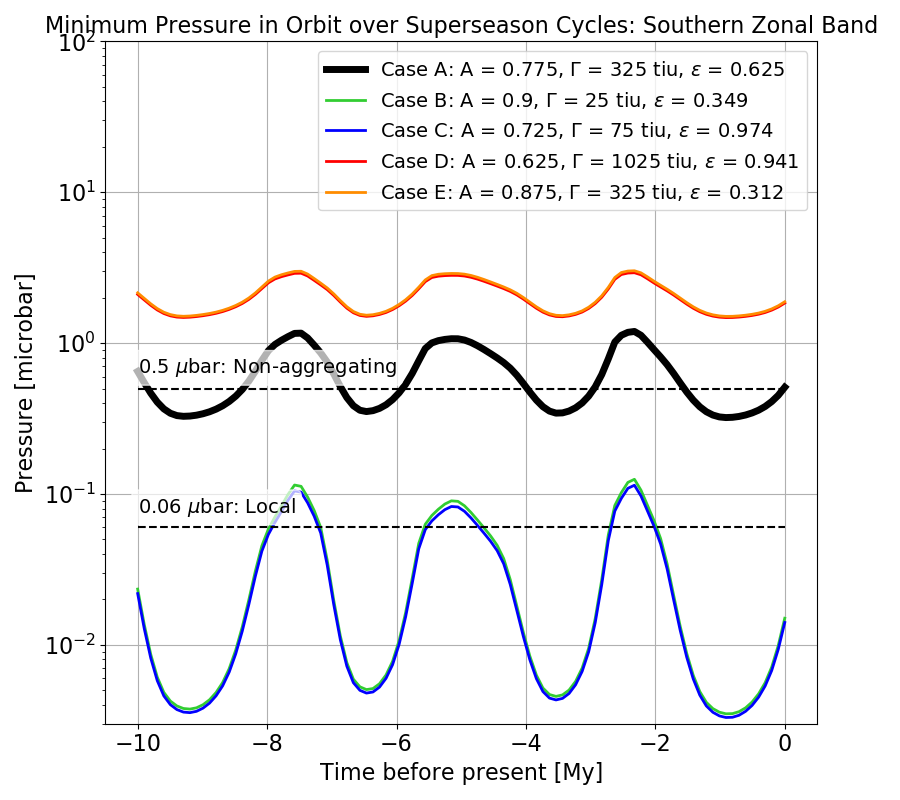}
		\caption{Annual minimum pressure experienced at Pluto's surface over the past 10 My for each of the five test cases, for the southern zonal band model (log scale). At this scale, Case B is nearly coincident with Case C, and Case D is nearly coincident with Case E.}
		\label{fig:minpress_band}
	\end{center}
\end{figure}

Five test cases are shown in Figure \ref{fig:press_band} on a linear scale, and in Figure \ref{fig:press_band_log} on a logarithmic scale to highlight the very low pressures near aphelion and northern winter solstice. All of the example cases have perihelion pressures of around 5 $\mu$bar, and then the pressure rapidly increases to 11.5 $\mu$bar in 2015. Compared to the reference model, the peak in the pressure curve is much sharper and the minimum is much broader, due to the lower thermal inertias. The pressure peak occurs earlier in the orbit, around 2016 rather than 2027 for the reference model. This is driven by the spatially-averaged insolation; it is highest near equinox (nearly concurrent with perihelion) when the southern zonal band of \N2 ice and SP are both being directly illuminated, and decreases as the subsolar latitude moves to the north after equinox and the zonal band moves into polar night. The extremely low pressures occur near aphelion and winter solstice, when the spatially-averaged insolation onto the \N2 ices is low, and are due in part to the low thermal inertias which allow for quick temperature and pressure changes.

Figure \ref{fig:minpress_band} shows the superseasonal behavior for the five test cases in the southern zonal band model. Three of the five cases predict minimum pressures below the haze aggregation limit. Two of those cases, B (green curve) and C (blue curve), predict a minimum pressure below the local atmospheric limit in nearly every orbit for the past 10 My. All of the test cases produce atmospheres that remain opaque to UV radiation throughout the past 10 My.



\subsection{Mobile Nitrogen Sensitivity Test}
A limitation of the results we present earlier in Section 3 is the use of a static \N2 distribution. On Pluto, atmospheric \N2 is able to condense onto previously \N2-free surfaces, a process not accounted for in a static model. We address this limitation here with a sensitivity study. The purpose of this study is to evaluate the effect mobile \N2 has on the minimum pressure over a Pluto orbit, consequently what effect it has on haze production, and under what conditions a static \N2 assumption is valid. A static \N2 distribution has two crucial benefits: (1) computation speed that allows us to explore a wide range of thermal parameters and long timescales and (2) an exact match to the observed \N2 distribution in 2015 (the only time a nearly-global map has been produced).

We simulated the effect of mobile \N2 by establishing the times and locations where the \N2 could condense onto the surface, by identifying when and where the substrate temperature is colder than the perennial \N2 ice's temperature. First, we calculated the bare substrate temperature as a function of time at each latitude, which depends on the substrate emissivity, albedo, and thermal inertia. We selected a value of 800 tiu for the thermal inertia, 0.9 for the emissivity, and investigate albedos between 0 and 1 in steps of 0.1 (similar to the values of 800 tiu for bare-ground thermal inertia and 1 for emissivity chosen in \citet{Bertrand2016}, \citet{Bertrand+2018}, \citet{Bertrand+2019}). The temperature at each location is also dependent on the incident diurnally-averaged solar insolation, which varies with latitude, time, and albedo. We began by using the Case A \N2 ice temperature from each of the four \N2 distributions presented in Sections 3.1 to 3.4. We assumed that regions of the surface that are colder than the \N2 ice at a given time will become covered by mobile \N2 ice via condensation from the atmosphere, forming a seasonal deposit. This includes regions of bare substrate and also regions with perennial \N2 deposits that have fractional abundances less than 1. For example, if a given area has perennial \N2 with a fractional abundance of 0.25, then the remaining 75\% of the area can become covered by seasonal \N2. The resultant seasonal \N2 ice appears with the same thermal parameters as the perennial \N2 and a fractional abundance of 0.5. Using the seasonal \N2 ice distribution as well as the perennial \N2 ice as prescribed by the static \N2 distribution, we recalculated the diurnally- and spatially-averaged insolation onto all of the \N2 ice as described in Section 2.2, and used it to recalculate the temperature and pressure behavior as a function of time using VT3D. If the model didn't predict a 2015 pressure in the range 11.5 $\pm$ 0.5 $\mu$bar, we adjusted the emissivity and recalculated the perennial-only temperatures, the resulting locations of seasonal deposits, and finally the pressure accounting for both perennial and seasonal deposits. This iterative process was repeated until the 2015 pressures fell in the range 11.5 $\pm$ 0.5 $\mu$bar. In general, the modified Case A including mobile \N2 for all four \N2 distributions and all substrate albedos required equivalent or lower emissivities than the corresponding unmodified perennial-only static Case A.

The following figures show the effect of seasonal \N2 on the pressure. We discuss the results in terms of three broad classes divided by substrate Bond albedo: $A_{sub}$ $<$ 0.3 (dark substrate), 0.3 $<$ $A_{sub}$ $<$ 0.7 (intermediate substrate), and $A_{sub}$ $>$ 0.7 (bright substrate). In 2015, regions of Pluto's surface with Bond albedos less than 0.3 include the dark maculae near the equator, while \N2-free regions with Bond albedos higher than 0.7 include the polar region north of 60\deg N and Eastern Tombaugh Regio \citep{Buratti+2017}. We discuss three types of pressure curves: (1) static, which use a static \N2 distribution and were presented earlier in Section 3, (2) mobile, P$_{2015}$-unconstrained, which have the same thermal parameters as the static cases but allow \N2 mobility causing the pressure in 2015 to be inconsistent with observations, and (3) mobile, P$_{2015}$-constrained, in which we include seasonal \N2 deposits and modify the emissivity to ensure the pressure in 2015 equals 11.5 $\mu$bar as described above. As a specific example, we first present the results from Reference Distribution Case A.

\begin{figure}
	\begin{center}
		\includegraphics[width =\textwidth]{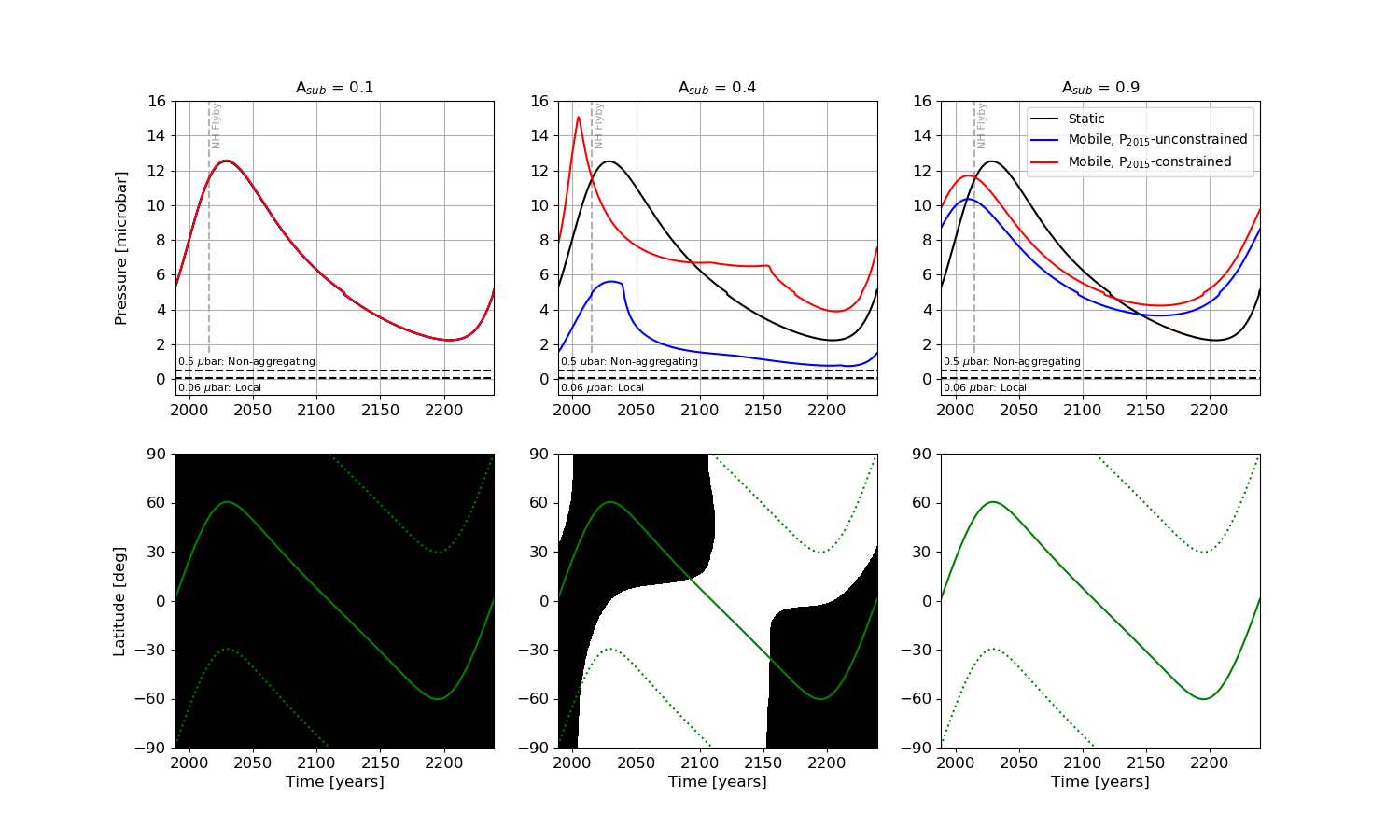}
		\caption{(top) Pressure vs. time for Reference Case A with three substrate albedos: $A_{sub}$ = 0.1 (left), 0.4 (middle), and 0.9 (right). The static Reference Case A pressure is shown in black, while the pressure from the mobile, P$_{2015}$-constrained case is shown in red. The blue line shows the mobile, P$_{2015}$-unconstrained case, which illustrates the need to adjust the emissivity in order for P$_{2015}$ to equal 11.5 μbar, especially for intermediate albedos. For the low $A_{sub}$ case (left), these three lines are the same. For intermediate $A_{sub}$ cases (middle), the mobile, P$_{2015}$-constrained case produces a sharper, higher, earlier peak in the pressure, and a higher minimum pressure. (bottom) Latitudes of seasonal \N2 deposits as a function of time for the mobile, P$_{2015}$-constrained Reference Case A with three substrate albedos: A$_{sub}$ = 0.1, 0.4, and 0.9. White indicates seasonal \N2 coverage at that latitude and time, while black indicates bare substrate. While many of these latitudes also contain perennial deposits (see Figure \ref{fig:n2}), only the seasonal deposits are indicated with white. The green solid line shows the subsolar latitude, while the green dotted line marks the edge of the polar night region: any latitudes polewards of the green dotted line experience no daylight. Between 2110 and 2150, the case with $A_{sub}$ = 0.4 predicts that Pluto would be fully covered in \N2, a ``Snowball Pluto'' scenario, while at other times there exists only a polar cap, at the north pole around the time of northern winter solstice (2195) and at the south pole near southern winter solstice (2029). When $A_{sub}$ = 0.1 (left), no seasonal deposits form at any latitudes and times, and when $A_{sub}$ = 0.9 seasonal deposits form at all latitudes and last for Pluto's entire orbit.}
		\label{fig:press_new_vol}
	\end{center}
\end{figure}

Using Reference Case A ($A_{N2}$ = 0.75, $\Gamma$ = 1225 tiu, $\epsilon$ = 0.593), dark substrate albedos ($A_{sub}$ $<$ 0.3) remain warmer than the \N2 ice temperature throughout much of the year. For example, for $A_{sub}$ = 0.1, the average substrate temperature is 44 K, while the average \N2 temperature is 36 K. Consequently, there are only small, short-lived seasonal \N2 deposits, or in the darkest albedo cases, no seasonal \N2 deposits at all, as seen in the left panels of Figure \ref{fig:press_new_vol}. The minimum pressure is not largely affected, remaining at 2 $\mu$bar, so the atmosphere remains haze-aggregating, global, and opaque to UV radiation. 

\begin{figure}
	\begin{center}
		\includegraphics[width =\textwidth]{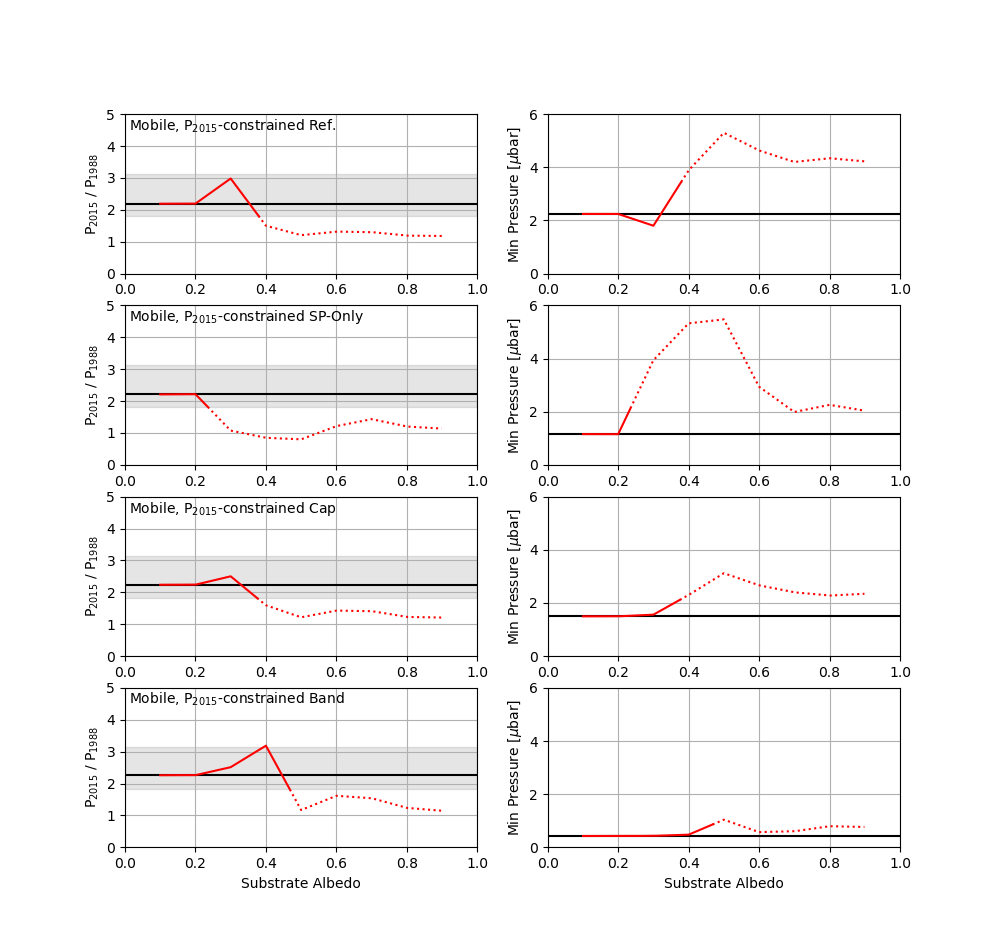}
		\caption{Results from the mobile \N2 ice sensitivity test. From top to bottom, they are: the Mobile, P$_{2015}$-constrained Reference Case A; Mobile, P$_{2015}$-constrained SP-Only Case A; Mobile, P$_{2015}$-constrained South Polar Cap Case A; and Mobile, P$_{2015}$-constrained Southern Zonal Band Case A. The left panels show show the effect of substrate albedo on the ratio of the 2015 pressure to the 1988 pressure, with the horizontal gray shading indicating the 3-$\sigma$ range from observations (1.82 - 3.14), while the right panels show the effect of substrate albedo on the minimum pressure over the course of an orbit. In all panels, the horizontal black line indicates the value from the respective static pressure curve, with no seasonal \N2. The transition from solid to dotted red lines occurs after the highest A$_{sub}$ with a valid pressure ratio.}
		\label{fig:pmin_pratio}
	\end{center}
\end{figure}

For bright substrate albedos ($A_{sub}$ $>$ 0.7), the substrate is typically colder than the \N2 ice temperature for most of the year. In Reference Case A, the \N2 albedo is 0.75, and when $A_{sub}$ $>$ $A_{N2}$ the substrate will absorb less sunlight than neighboring \N2 ice deposits at the same latitude, tending to make the substrate colder than \N2 ice. For example, with $A_{sub}$ = 0.9, the average substrate temperature is 26 K, compared to 36 K for the \N2 ice temperature. This results in a ``Snowball Pluto'' scenario, in which the entire surface is covered by \N2 ice deposits for Pluto's entire orbit, as shown in the right panels of Figure \ref{fig:press_new_vol}. Observations of Pluto's surface composition are inconsistent with a surface entirely covered by \N2 ice in 2015 \citep{Schmitt+2017, Protopapa+2017, Protopapa+2020}. The bright substrate cases have the peak in the pressure occurring earlier within Pluto's orbit, closer to perihelion, and the variation in the pressure over the orbit is reduced. Consequently, the ratio of the predicted 2015 to 1988 pressures is below the allowable 3-$\sigma$ range and the predicted atmosphere is not doubling (or tripling) as Pluto's atmosphere was observed to do in that time period (see the left panels of Figure \ref{fig:pmin_pratio}). We would not consider these bright substrate cases to be acceptable models based on this constraint and the surface composition observations.

For intermediate substrate albedos (0.3 $<$ $A_{sub}$ $<$ 0.7), seasonal \N2 deposits condense onto and subsequently sublime away from the substrate at various latitudes and times. Figure \ref{fig:press_new_vol} shows an example of this, for a substrate albedo of 0.4. The top panel shows three sets of pressure vs time: (1) the black line shows the static case A with no seasonal \N2, the same as what is shown in Figure 5, (2) the blue line shows the mobile, P$_{2015}$-unconstrained case A and (3) the red curve shows the mobile, P$_{2015}$-constrained case A. The blue curve demonstrates the need to adjust the emissivity once mobile \N2 is added; the pressure is uniformly lower for the mobile, P$_{2015}$-unconstrained case than for the mobile, P$_{2015}$-constrained case (red). Decreasing the emissivity from 0.593 to 0.395 increases the pressure in 2015 to 11.5 $\mu$bar in order to be consistent with the observed pressure. The mobile, P$_{2015}$-constrained case has a minimum pressure of 4 $\mu$bar and a ratio P$_{2015}$/P$_{1988}$ of 1.5, which is shown in Figure \ref{fig:pmin_pratio}. The bottom panel of Figure \ref{fig:press_new_vol} shows the latitudes where seasonal \N2 deposits form for the mobile, P$_{2015}$-constrained case A. While many of these latitudes also contain perennial deposits (see Figure \ref{fig:n2}), only the seasonal deposits are indicated in Figure \ref{fig:press_new_vol}. Between 2110 and 2150, this case predicts that Pluto would be fully covered in \N2, a ``Snowball Pluto'' scenario, while at other times there exists only a polar cap: at the north pole around the time of northern winter solstice (2195) and at the south pole near southern winter solstice (2029). However, for this case, and almost all of the other intermediate albedo cases from the four distributions, the predicted ratio of pressures in 2015 and 1988 is below the allowed 3-$\sigma$ range. The top panel of Figure \ref{fig:press_new_vol} shows the predicted pressure vs. time has a sharper, higher, and earlier peak in addition to the higher minimum. The peak occurs prior to 2015, so while the atmosphere does double in pressure soon after 1988, the pressure is already decreasing at the time of the New Horizons flyby and the ratio of pressures in 2015 and 1988 is only 1.5. The overturning of pressure prior to 2016 is inconsistent with observations, which show a monotonic increase in pressure through 2016 \citep{Meza+2019}. The early peak and inconsistent P$_{2015}$/P$_{1988}$ ratio were predicted in most of the intermediate albedo cases for the four distributions.

While the previous discussion focused on the reference distribution case A, the general behavior was repeated in our analysis of the other \N2 distributions as well. Figure \ref{fig:pmin_pratio} shows the minimum pressure and ratio of 2015 to 1988 pressures as a function of substrate albedo for Case A for each of the four distributions. For substrate albedos less than 0.3, there is little to no seasonal \N2, so the minimum pressure, and consequently the time spent below any of the haze limiting pressures, do not differ greatly from the static cases. For intermediate albedos, seasonal \N2 deposits condense and sublime away at various times. In general, the inclusion of seasonal deposits increases the minimum pressure. For the Southern Zonal Band Case A specifically, the minimum pressure is increased above the haze aggregation limit when $A_{sub}$ $>$ 0.5, so the inclusion of seasonal \N2 deposits prevents the atmosphere from becoming non-aggregating. However, for all distributions, cases with $A_{sub}$ $\geq$ $\sim$0.3 are inconsistent with observations. The peak in pressure occurs too soon after perihelion, contradicting observations of a monotonic increase in surface pressure through 2016 (Meza et al. 2019) and also failing to reproduce the two- to three-fold increase in surface pressure observed between 1988 and 2015 (see the left panels of Figure \ref{fig:pmin_pratio}). For bright substrate albedos ($A_{sub}$ $>$ 0.7), seasonal \N2 deposits would cover Pluto completely, which is inconsistent with surface composition measurements \citep{Schmitt+2017,Protopapa+2017,Protopapa+2020}.

Seasonal \N2 deposits could have different thermal properties than their perennial counterparts, unlike the assumption used here that perennial and seasonal deposits share the same albedo, thermal inertia, and emissivity. Freshly condensed \N2 ice could be brighter than older \N2 ice, which has been processed by incident radiation and could contain contaminants that have fallen onto it. Fresh \N2 ice could be transparent, revealing the albedo of the substrate below \citep{Eluszkiewicz1991}; the substrate albedo should then be used in energy balance calculations. If the substrate albedo is less than the perennial \N2 ice's albedo, then the effect of transparent seasonal deposits is to warm the \N2 ice and increase the surface pressure (unless the emissivity is also increased in order to keep P$_{2015}$ at 11.5 $\mu$bar). If the substrate albedo is greater than the perennial \N2 ice's albedo, then transparent seasonal deposits create a cooler, lower pressure atmosphere, compared to the situation in which the seasonal and perennial deposits have the same albedo. Seasonal \N2 deposits will be thinner than perennial deposits which have been built up over many Pluto years. Annual condensation rates were calculated to be on the order of a centimeter by \citet{Bertrand+2018}. It may therefore be more accurate to use the substrate thermal inertia in energy balance calculations, since the material contained within one thermal skin depth (5 - 50m) will be primarily substrate with just a thin coating of \N2. Freshly condensed \N2 might also have lower emissivity than older \N2 deposits, owing to the smaller grains (older \N2 could sinter together over time into larger grains) \citep{Stansberry+1996}. Exploring all of these possibilities introduces considerable complexities to the problem of modeling Pluto's pressure.

What we have shown here is not all-encompassing; we have not explored the full range of thermal parameters for the seasonal \N2 ice, and we have not iterated this process to see if the newly recalculated \N2 temperature behavior reproduces the same seasonal \N2 locations. Including mobile \N2 introduces a plethora of new parameters space to explore (substrate thermal properties, seasonal \N2 deposit thermal properties and fractional abundance), which add complexity to the problem of matching the New Horizons observations, such as the pressure and the nitrogen distribution in 2015. Exploring all of these possibilities is outside the scope of this paper. By providing this sensitivity section, we hope to motivate our assumption of static, perennial \N2. For seasonal \N2 deposits with the same properties as their perennial counterparts, we found no cases which produced widespread seasonal deposits and simultaneously matched observations of the 2015 pressure and reproduced the two- to three-fold increase in pressure between 1988 and 2015. For substrate albedos less than 0.3, the substrate is too warm for \N2 condensation. For intermediate substrate albedos, the predicted pressure variations between 1988 and 2016 are inconsistent with observations from occultations. For bright substrate albedos, the resulting surface composition and the pressure behavior are both inconsistent with observations and can thus be ruled out. 

From this sensitivity study we conclude the following:
\begin{enumerate}
	\item The inclusion of mobile \N2 tends to raise the minimum pressure, because the emissivity must be decreased in order to match the observed 2015 pressure. This restricts the section of parameter space that leads to haze disruption or atmospheric collapse.
	\item There is likely not a large amount of \N2 in the unobserved southern hemisphere. Most cases that have expansive and/or long-lived seasonal deposits in the southern hemisphere failed to match observations of the change in surface pressure between 1988 and the present. We concluded this from our static-\N2 models as well; the South Polar Cap and Southern Zonal Band distributions required very low \N2 fractional abundances (less than 0.2) to match observations. Meza et al. 2019 reached a similar conclusion from their modeling results as well.
	\item Our conclusions about haze interruption based on static \N2 distributions are not likely to be changed based on the inclusion of mobile \N2.
\end{enumerate}

\section{Haze Implications}

In Pluto's current orbit, our reference model does not produce any case where the pressure drops low enough to interrupt haze. There are no combinations of parameters, namely Bond albedo, thermal inertia, and emissivity, which are simultaneously capable of reproducing the observed 2015 flyby pressure and having a minimum pressure below any of the haze-disruption pressures, in the current orbit. Additionally, on long timescales, none of the five test cases in our reference model produce pressures that fall below the haze-disruption pressures. The modeled atmosphere remains haze-aggregating, global, and opaque to UV radiation during the 10 My period we investigated.

Southern \N2 is necessary for haze to be interrupted. Our south polar cap and southern zonal band models both predict that haze aggregation could stop at some point during the orbit, although in the case of our polar cap this is only possible for special orbital configurations when northern summer solstice and perihelion occur at the same time, and then only for low-albedo, low-thermal inertia cases. In the case of the southern zonal band model, haze aggregation is stopped between aphelion and northern winter solstice in Pluto's current orbit and in many past orbits going back 10 My, for most cases in the allowed parameter space. Stopping haze aggregation for a portion of the orbit could cause the appearance and size of the haze particles being deposited to vary seasonally. The haze was observed globally at the time of the New Horizons flyby, but it was brighter towards the north, probably indicating greater haze mass \citep{Cheng+2017}. Deposition rates could be dependent on the brightness, which could vary seasonally. Thus, locations on the surface with a higher deposition rate could be covered with more monomer haze particles than others, explaining the heterogeneity. As demonstrated in \citet{Bertrand+2017}, haze production rates as a function of latitude and time can be determined based on the assumed UV flux at the top of Pluto's atmosphere and the opacity of the atmosphere. This same technique could be applied to our results, in order to determine which latitudes would experience the largest decrease in haze production resulting from atmospheric collapse. If meridional circulation is weak, these latitudes would also experience the largest decrease in haze deposition.

The atmosphere resulting from the southern zonal band model becomes local between aphelion and northern winter solstice, but only for the lowest thermal inertias. When the atmosphere becomes local, the sublimation winds are equal in magnitude to the atmosphere's sound speed, and thus there will be large pressure variations across the surface \citep{TraftonStern1983}. As a result, the atmosphere becomes patchy and Io-like, extending only over the warmest patches of the surface. Any haze deposition would be restricted to these patches, which could build up surface contrasts. It could also reinforce existing contrasts. All else being equal, the darkest \N2 surfaces will be the warmest and could maintain an atmosphere above them. If the deposition of haze particles darkens the surface further, it would create a positive feedback that enhances existing surface contrasts. Conversely, a local atmosphere could shield the underlying surface from UV light, preventing ice-phase photolysis. Whether this would lead to positive or negative feedback depends on the relative albedo of the gas-phase and ice-phase photolysis products, and their rates of production.


A complication we have not considered here is a time-variable CH$_4$ mixing ratio in the atmosphere. The pressures we investigate here as being relevant to haze production (0.5 $\mu$bar haze aggregation limit, 0.06 $\mu$bar local atmosphere limit, and the 10$^{-3}$ to 10$^{-4}$ $\mu$bar atmospheric transparency limit) are determined from the atmospheric structure as observed in 2015 by New Horizons. Over time however, the mixing ratio of CH$_4$ could vary, changing the altitude at which the photochemical reactions producing the haze occur. A variable CH$_4$ mixing ratio would have implications for haze chemistry, changing the color and composition, as well as the production rate. For example, if the mixing ratio was about 10$^{-3}$ times less than it is currently, the atmospheric transparency limit would be 10$^3$ times higher, at about 1 $\mu$bar, and many of our cases would become transparent to UV radiation. However, models with variable mixing ratios of CH$_4$ show much less variation than that, on the order of a factor of 10 to 20 \citep{Bertrand+2019}, and the CH$_4$ mixing ratio tends to be higher when the surface pressure is lowest \citep{Bertrand2016}. This suggests that, while a variable CH$_4$ mixing ratio may effect the photochemical products in Pluto's atmosphere, it does not lead to Pluto's atmosphere becoming UV-transparent. 

\citet{Grundy+2018} and \citet{Bertrand+2019b} describe other methods that could explain the observed surface heterogeneity, which we briefly summarize here. One mechanism could be differing thermal processing of the haze particles as they settle through the atmosphere, perhaps due to latitudinal or seasonal changes in the amount or type of hydrocarbons available to stick onto the haze monomers. If the haze particles are not all uniform but instead follow a distribution of characteristics such as size or albedo, then different parts of the distribution could respond differently in various surface environments. Another possible mechanism is cyclical burial and exhuming of haze particles, where the different surface appearances could represent freshly fallen hazes versus exhumed, previously buried haze particles. Over SP, katabatic winds blowing downslope could concentrate haze particles on the ice sheet, counteracting the sublimation winds' tendency to blow haze particles off of it \citep{Bertrand+2019b}; aeolian processes could be important at the locations of other \N2 deposits as well. \citet{Protopapa+2020} suggest that a single coloring agent, very similar to the Titan-like tholin of \citet{Khare+1984}, can account for all of Pluto's colors (from red to yellow). They suggest that Pluto's coloration is the result of photochemical products mostly produced in the atmosphere, concurring with \citet{Grundy+2018}. Variations in color are to be attributed to variations in abundance and grain size of the haze particles.

\section{Conclusions}

\begin{table*}[]
	\caption{Summary of the haze disruption results for each of the spatial \N2 distributions we investigate.}
	\begin{center}
	\footnotesize
	\setlength\tabcolsep{1pt}
	\begin{tabular}{|c|c|c|c|c|}
		\hline
		\multicolumn{2}{|c|}{\multirow{2}{*}{}}                      & \textbf{Non-Aggregating} & \textbf{Local} & \textbf{UV-Transparent} \\ \cline{3-5} 
		\multicolumn{2}{|c|}{}                                       & $<$0.5 $\mu$bar            & $<$0.06 $\mu$bar     & $<$10$^{-3}$ to $<$10$^{-4}$$\mu$bar                 \\ \hline\hline
		\multirow{2}{*}{\textbf{Reference Model}}     & Current      &        -              &         -       &            -           \\ \cline{2-5} 
		& Superseasons &          -            &        -        &          -             \\ \hline
		\multirow{2}{*}{\textbf{Sputnik Planitia - Only}}             & Current      & possible                    &  -              &      -                 \\ \cline{2-5} 
		& Superseasons & possible                    & possible           &              -         \\ \hline
		\multirow{2}{*}{\textbf{South Polar Cap}}     & Current      &           -           &        -        &        -               \\ \cline{2-5} 
		& Superseasons & possible                    &          -      &               -        \\ \hline
		\multirow{2}{*}{\textbf{Southern Zonal Band}} & Current      & probable                    & possible              &          -             \\ \cline{2-5} 
		& Superseasons & probable                  & possible            &            -           \\ \hline
	\end{tabular}
	\label{table:summary}
	\end{center}
\end{table*}

Table \ref{table:summary} summarizes the results for each of the four \N2 distributions we investigate here, for Pluto's current orbit and configurations experienced over the past 10 My. `Possible' indicates that a particular model predicts pressures for 1 or 2 of the test cases indicative of an atmosphere with the given characteristic (non-aggregating, local or UV transparent) for some portion of the orbit, while `probable' indicates that 3 or more of the test cases predicted atmospheres with that characteristic. Table \ref{table:summary} is based on the five test cases for each \N2 distribution, but since the test cases were chosen to span the allowed parameter ranges, they are indicative of the whole parameter space. For the reference model, which has a bare southern hemisphere, haze production is not predicted to be interrupted at all, and the atmosphere will not collapse, neither in the current orbit nor over the past 10 My. Non-aggregating and local atmospheres are possible using the SP-Only distribution, but this distribution was included for comparison purposes only and is not realistic. Therefore, southern \N2 in some form is required to produce pressures below any of the haze-disruption pressures we considered. We investigated two example southern \N2 distributions: a south polar cap extending from the pole to 60\deg S with a fractional abundance of 20\% and a southern zonal band between 35\deg S and 55\deg S, also with a fractional abundance of 20\%. Other southern distributions are of course possible, but we chose these two to be representative of some of the possibilities. Atmospheric collapse, when the pressure becomes too low to support a global atmosphere, only occurs in our southern zonal band model, and only for low thermal inertias ($<$200 tiu). Across all realistic \N2 distributions and cases considered here, the minimum pressure is predicted to be between 0.01 - 3 $\mu$bar in Pluto's current orbit, and the pressure has remained above 0.004 $\mu$bar over the past 10 My.

In general, the \N2 ices on the surface collectively re-radiate the insolation absorbed by only the illuminated ices. If more ice coverage is added to the southern hemisphere, currently in polar night, then these unilluminated ices will not absorb solar energy, but they will emit energy. Thus, the presence of obscured southern \N2 ices can lower the minimum pressure experienced over an orbit. However, in order to satisfy the constraints (doubling of the surface pressure since 1988 and an 11.5 $\mu$bar pressure in 2015), we found that \N2 distributions including southern \N2 required much lower thermal inertias. From the mobile \N2 sensitivity study, we concluded that the addition of mobile \N2 tends to increase the minimum pressure experienced during Pluto's orbit, because the emissivity of the \N2 ice must be decreased in order to match the observed pressure in 2015. Thus, \N2 mobility decreases the likelihood of haze disruption or atmospheric collapse.


The amount of the southern hemisphere that is obscured in polar night will not decrease until after solstice occurs in 2029, and the entire southern hemisphere won't be visible until equinox occurs 100 years after that. The southern hemisphere could be thermally mapped when it is in polar night, providing a means to determine the spatial distribution of \N2 in the near future rather than a century from now. Our model predicts that there can only be small perennial southern deposits, since we were unable to match observable constraints for southern zonal bands or south polar caps with fractional abundances above 20\%, or large, seasonal southern deposits. Another possible scenario is that the southern deposits could have different properties (e.g. a larger deposit with a lower fractional abundance, or different thermal properties) than the northern deposits, which we have not explored here. 

Recent analysis of ground-based stellar occultations report a monotonic increase in Pluto's pressure between 1988 and 2016 \citep{Meza+2019}. This is consistent with nearly all of our static models, with the exception of test case C from the southern zonal band distribution, which predicts a turnover in pressure in 2008 and test case B from the southern zonal band distribution, which is approximately flat in pressure in the 2010s. \citet{Arimatsu+2020} observed a single-chord occultation in 2019 which showed a large drop in surface pressure, from 11.5 $\mu$bar in 2015 to 9.56$^{+0.52}_{-0.34}$ $\mu$bar in 2019. This pressure decrease is marginally significant at the 2.4-$\sigma$ level; if real, it represents an earlier and more rapid decrease than all of the models we present here, with the exception of the Southern Zonal Band cases B and C (however, the pressure peak in case B occurs in 2008, inconsistent with \citet{Meza+2019} occultation results). All of our models predict a turnover in the pressure by the 2030s, when the surface pressure will begin to decrease as Pluto moves toward aphelion and the subsolar latitude retreats to the southern hemisphere. However, the date of the turnover and the rate of the decline in pressure varies between distribution and chosen parameters in our model. Observations of the atmospheric pressure in the next few decades will thus be crucial for determining which \N2 distributions and which (\textit{A},$\Gamma$,$\epsilon$) triplets best represent Pluto.

\section{Acknowledgments}
This work was supported in part by NASA ROSES SSW grant NNX15AH35G and by NASA's New Horizons mission to the Pluto system. The authors would like to thank Tanguy Bertrand for illuminating discussions and feedback, as well as James Keane, Amanda Sickafoose, and Anja Genade for reviewing early versions of the appendix. We would also like to thank two anonymous reviewers, for greatly improving this paper with their feedback.



\clearpage
\appendix
\section{Approximating Temperatures with VT3D}
\subsection*{Temperature from Analytic Approximation}
Volatile Transport 3D (VT3D) uses an analytic approximation of the temperature evolution as an initial solution for the more accurate numerical solution. On its own, the analytic solution is often a good approximation and it is computationally more expedient. This appendix explains how to use the analytic approximation to calculate surface pressures over a period of one Pluto orbit, using the reference model as described in the paper. 

The diurnally- and spatially-averaged incident insolation $S(t)$ can be represented using an analytic Fourier approximation:
\begin{eqnarray}
S_0 &=& \frac{1}{P}\int_0^P S(t)dt \\ 
S_m &=& \frac{2}{P}\int_0^P S(t)e^{-im\omega t}dt \quad \quad m > 0
\end{eqnarray}
where P is the period of the solar forcing (in this case, one Pluto year) and $\omega$ is the corresponding frequency. $m$ is an integer corresponding to the $m$th Fourier term. For the reference model insolation, the first 11 Fourier terms are provided in Table \ref{table:sm}. These terms are for the diurnally- and spatially-averaged insolation onto the N$_2$-covered regions. The diurnally-averaged incident insolation as a function of latitude $\lambda$ can be calculated via:

\begin{equation}
\overline{S(\lambda,t)} = \frac{\sin{\lambda}\sin{\lambda_0}h_{max}+\cos{\lambda}\cos{\lambda_0}\sin{h_{max}}}{\pi}\frac{S_{1AU}}{r^2}
\end{equation}
where $\lambda_0$ is the subsolar latitude, $S_{1AU}$ = 1361 W m$^{-2}$, and $r$ is the heliocentric distance, in AU. The maximum illuminated hour angle at that latitude, $h_{max}$, can be found using: $\cos h_{max} = max( −1, min(-\tan\lambda\tan\lambda_0, 1))$. The time variable, $t$, represent time within in one Pluto year, and timesteps must be larger than one Pluto day (we used $\Delta t$ = 0.5 Earth years). To spatially-average over the N$_2$-covered regions, we calculate:
\begin{equation}
S(t) = \frac{\int_{N_2}\overline{S(\lambda,t)}\Omega d\Omega}{\int_{N_2}\Omega d\Omega}
\end{equation}
where $\Omega$ is the solid angle area of a patch on the surface covered by N$_2$ and the integral is performed over all patches.

\setcounter{table}{0}
\begin{table}[h]
	\caption{Fourier terms for the incident insolation for the reference model.}
	\centering
	\begin{tabular}{||c | c||} 
		\hline
		m & $S_m$ $[W/m^2]$ \\ [0.5ex] 
		\hline\hline
		0 & 0.220561 \\ 
		1 & 0.115454 - 0.136762i \\ 
		2 & 0.043688 - 0.068281i \\ 
		3 & 0.015757 - 0.029367i \\ 
		4 & 0.007107 - 0.011570i \\ 
		5 & 0.003849 - 0.004378i \\ 
		6 & 0.002244 - 0.001651i \\ 
		7 & 0.001404 - 0.000616i \\ 
		8 & 0.000920 - 0.000234i \\ 
		9 & 0.000615 - 0.000097i \\ 
		10 & 0.000408 - 0.000062i \\ 	[1ex] 
		\hline
	\end{tabular}
	\label{table:sm}
\end{table}

These insolation terms can be converted into temperatures using the following equation:
\begin{equation}
T(\zeta,t) = -\frac{F\zeta}{\Gamma\sqrt{\omega}} + T_0 + Re\left[\sum_{m=1}^MT_me^{im\omega t}e^{\sqrt{im}\zeta}\right] \label{tm}
\end{equation}
$T_0$ is the average temperature assuming thermal emission balances solar insolation and internal heat flux, $F$: $T_0 = ([(1-A)S_0+F]/\epsilon\sigma)^{1/4}$. $\zeta = z/Z$ is the unitless depth of the layer, scaled by the skin depth, $Z = \sqrt{k/(\rho c \omega)}$. For N$_2$ ice, we use density $\rho$ = 1000 kg m$^{-3}$, specific heat c = 1300 \textit{J kg$^{-1}$ K$^{-1}$}, and calculate the heat conductivity $k$ based on the selected thermal inertia value ($\Gamma = \sqrt{k\rho c}$). For surface temperatures, the depth \textit{z} = 0.

Each temperature Fourier coefficient for $m > 0$ is given by:
\begin{equation}
T_m = \frac{(1-A)S_m}{\Phi_E(T_0)}\frac{4}{4+\sqrt{im}\Theta_S(T_0) + im\Theta_A(T_0)} \quad \quad m > 0
\end{equation}
where $\Phi_E$ is the derivative of the thermal emission with respect to temperature:
\begin{equation}
\Phi_E(T_0) = 4\epsilon\sigma T_0^3
\end{equation}
where the Stefan-Boltzmann constant $\sigma$ = 5.67 x 10$^{-8}$ W m$^{-2}$. The dimensionless thermal parameters $\Theta_S$ (buffering of volatile temperature due to thermal conduction to neighboring layers) and $\Theta_A$ (buffering due to latent heat of sublimation) are defined as:
\begin{equation}
\Theta_S(T_0) = \frac{\sqrt{\omega}\Gamma}{\Phi_E(T_0)/4}
\end{equation}
\begin{equation}
\Theta_A(T_0) = \frac{\omega\frac{L_s}{f_vg}\frac{dp_s}{dT_V}\bigg\rvert_{T_0}}{\Phi_E(T_0)/4}
\end{equation}
where $L_s$ is the latent heat of sublimation for N$_2$: approximately 2.7 x 10$^5$ J kg$^{-1}$ for $\alpha$-phase (below 35.6 K) and 2.4 x 10$^5$ J kg$^{-1}$ for $\beta$-phase (above 35.6 K). The surface gravity $g$ is 0.62 m s$^{-2}$. The fraction of the surface covered by nitrogen ice, using our reference map, $f_v$, is 0.102. $dp_s/dT_V$ is the derivative of the vapor pressure with respect to the volatile temperature, evaluated at $T_0$:
\begin{equation}
\frac{dp_s}{dT_V}\bigg\rvert_{T_0} = \frac{L_sm_Vp_{s}(T_0)}{k_BT_0^2}
\end{equation}
where $m_v$ is the molecular mass of N$_2$, $p_s(T_0)$ is the equilibrium vapor pressure above solid N$_2$ at temperature $T_0$, and $k_B$ is the Boltzmann constant. 

\subsection*{Selecting \textit{A}, $\Gamma$, and $\epsilon$}
As described in this paper, VT3D has three free parameters that describe the nitrogen frost: the Bond albedo, \textit{A}, the thermal inertia, $\Gamma$ (in units of ``tiu'', J m$^{-2}$ K$^{-1}$ s$^{-1/2}$), and the emissivity, $\epsilon$. We select values for \textit{A} and $\Gamma$, and then choose a corresponding value for $\epsilon$ such that the pressure predicted by the model at the time of the New Horizons flyby is 11.5 $\mu$bar. It the paper, we iteratively calculate pressures with different emissivities until we find a solution that predicts the correct pressure in 2015. Here, we present a polynomial fit to the relationship this process derived. The coefficients $k_i$ (which are each a function of \textit{A}) in Table \ref{table:k} can be used along with the equation below to calculate the emissivity needed for the chosen albedo and thermal inertia value. The relationship predicts the necessary emissivity to within 2\% of the correct value for most \textit{A} and $\Gamma$ values. Once the emissivity value for the chosen \textit{A} and $\Gamma$ has been calculated, Equation \ref{tm} can be used to calculate the temperature at every point \textit{t} within Pluto's orbit.

\begin{equation}
\epsilon(A,\Gamma) = k_0(A) + k_1(A)\Gamma + k_2(A)\Gamma^2 + k_3(A)\Gamma^3 + k_4(A)\Gamma^4
\end{equation}

\begin{table}[]
	\caption{Coefficients (as a function of albedo) needed to calculate the emissivity.}
	\centering
	\begin{tabular}{||c|c|c|c|c|c||}
		\hline
		\textbf{Albedo} & \textbf{k$_0$} & \textbf{k$_1$} & \textbf{k$_2$} & \textbf{k$_3$} & \textbf{k$_4$} \\ \hline\hline
		0.500           & 2.378e+00     & -2.141e-03    & 1.760e-06     & -7.190e-10    & 1.150e-13     \\ \hline
		0.525           & 2.254e+00     & -2.116e-03    & 1.804e-06     & -7.581e-10    & 1.240e-13     \\ \hline
		0.550           & 2.128e+00     & -2.081e-03    & 1.835e-06     & -7.910e-10    & 1.319e-13     \\ \hline
		0.575           & 2.001e+00     & -2.033e-03    & 1.851e-06     & -8.167e-10    & 1.385e-13     \\ \hline
		0.600           & 1.872e+00     & -1.974e-03    & 1.855e-06     & -8.378e-10    & 1.447e-13     \\ \hline
		0.625           & 1.741e+00     & -1.901e-03    & 1.840e-06     & -8.488e-10    & 1.488e-13     \\ \hline
		0.650           & 1.607e+00     & -1.808e-03    & 1.795e-06     & -8.429e-10    & 1.496e-13     \\ \hline
		0.675           & 1.473e+00     & -1.700e-03    & 1.731e-06     & -8.268e-10    & 1.484e-13     \\ \hline
		0.700           & 1.336e+00     & -1.575e-03    & 1.642e-06     & -7.961e-10    & 1.444e-13     \\ \hline
		0.725           & 1.198e+00     & -1.430e-03    & 1.521e-06     & -7.467e-10    & 1.365e-13     \\ \hline
		0.750           & 1.060e+00     & -1.272e-03    & 1.378e-06     & -6.843e-10    & 1.260e-13     \\ \hline
		0.775           & 9.222e-01     & -1.097e-03    & 1.206e-06     & -6.034e-10    & 1.115e-13     \\ \hline
		0.800           & 7.870e-01     & -9.185e-04    & 1.026e-06     & -5.191e-10    & 9.667e-14     \\ \hline
		0.825           & 6.551e-01     & -7.360e-04    & 8.343e-07     & -4.262e-10    & 7.994e-14     \\ \hline
		0.850           & 5.288e-01     & -5.562e-04    & 6.371e-07     & -3.277e-10    & 6.177e-14     \\ \hline
		0.875           & 4.102e-01     & -3.872e-04    & 4.459e-07     & -2.302e-10    & 4.352e-14     \\ \hline
		0.900           & 3.022e-01     & -2.428e-04    & 2.810e-07     & -1.458e-10    & 2.766e-14     \\ \hline
		0.925           & 2.064e-01     & -1.265e-04    & 1.449e-07     & -7.470e-11    & 1.412e-14     \\ \hline
		0.950           & 1.247e-01     & -4.950e-05    & 5.585e-08     & -2.858e-11    & 5.377e-15     \\ \hline
		0.975           & 5.706e-02     & -1.293e-05    & 1.536e-08     & -8.019e-12    & 1.512e-15     \\ \hline
	\end{tabular}
	\label{table:k}
\end{table}


\bibliographystyle{elsarticle-harv} 
\bibliography{master}





\end{document}